\documentclass[pdftex,twocolumn,epjc3]{svjour3}          

\RequirePackage[T1]{fontenc}

\smartqed  

\RequirePackage{graphicx}
\RequirePackage{mathptmx}      
\RequirePackage{flushend}
\RequirePackage[numbers,sort&compress]{natbib}
\RequirePackage[colorlinks,citecolor=blue,urlcolor=blue,linkcolor=blue]{hyperref}

\usepackage{booktabs, url}
\usepackage{longtable, tabularx, multirow, amsmath}
\usepackage[table,xcdraw]{xcolor}
\usepackage{pdflscape}
\usepackage[export]{adjustbox}
\usepackage{array}
\usepackage{supertabular}
\usepackage{subcaption}
\usepackage{pbox}
\usepackage{colortbl}
\usepackage{subcaption}
\usepackage{comment}
\usepackage[title]{appendix}
\usepackage{stfloats}

\usepackage{hyperref}

\hypersetup{
    colorlinks=true,
    linkcolor=blue,
    filecolor=blue,      
    urlcolor=blue,
    citecolor=blue
}

\newcolumntype{P}[1]{>{\centering\arraybackslash}p{#1}}
\newcolumntype{M}[1]{>{\centering\arraybackslash}m{#1}}

\journalname{Eur. Phys. J. C}

\begin{document}

\title{A Systematic Literature Review on Wearable Health Data Publishing under Differential Privacy}

\author{
        Munshi Saifuzzaman\thanksref{e1, addr1} \href{https://orcid.org/0000-0002-9236-5554}{\includegraphics[width=.125in,height=.125in,clip,keepaspectratio]{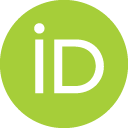}}
        \and
        Tajkia Nuri Ananna\thanksref{e1, addr1} \href{https://orcid.org/0000-0001-7385-980X}{\includegraphics[width=.125in,height=.125in,clip,keepaspectratio]{Images/ORCID.png}}
        \and
        Mohammad Jabed Morshed Chowdhury\thanksref{e2, addr2} \href{https://orcid.org/0000-0003-4476-8882}{\includegraphics[width=.125in,height=.125in,clip,keepaspectratio]{Images/ORCID.png}}
        \and
        Md Sadek Ferdous\thanksref{e3, addr3} \href{https://orcid.org/0000-0002-8361-4870}{\includegraphics[width=.125in,height=.125in,clip,keepaspectratio]{Images/ORCID.png}}
        \and
        Farida Chowdhury\thanksref{e4, addr1} \href{https://orcid.org/0000-0001-9902-6291}{\includegraphics[width=.125in,height=.125in,clip,keepaspectratio]{Images/ORCID.png}}
}

\thankstext{e1}{E-mail: \{munshi94, tajkia92\}@student.sust.edu}
\thankstext{e2}{E-mail: M.Chowdhury@latrobe.edu.au}
\thankstext{e3}{E-mail: sadek.ferdous@brac.ac.bd}
\thankstext{e4}{E-mail: farida-cse@sust.edu}

\institute{
        Shahjalal University of Science and Technology, Kumargaon, Sylhet-3114, Bangladesh\label{addr1}
        \and
        BRAC University, Dhaka-1212, Bangladesh\label{addr3}
        \and
        La Trobe University, Bundoora, VIC, 3086, Australia\label{addr2}
}

\date{Received: date / Accepted: date}

\twocolumn[
\maketitle
\begin{@twocolumnfalse}

\begin{abstract}
Wearable devices generate different types of physiological data about the individuals. These data  can  provide  valuable  insights  for  medical researchers and clinicians that  cannot be availed  through  traditional  measures. Researchers   have   historically  relied  on  survey  responses  or  observed  behavior. Interestingly, physiological data can provide a richer amount of user cognition than that obtained from any  other  sources,  including  the  user  himself. Therefore, the inexpensive consumer-grade wearable devices have become a point of interest for the health researchers. In addition, they are also used in continuous remote health monitoring and sometimes by the insurance companies. However, the biggest concern for such kind of use cases is the privacy of the individuals. There are a few privacy mechanisms, such as abstraction and \textit{k}-anonymity, are widely used in information systems. Recently, Differential Privacy (DP) has emerged as a proficient technique to publish privacy sensitive data, including data from wearable devices. In this paper, we have conducted a Systematic Literature Review (SLR) to identify, select and critically appraise researches in DP as well as to understand different techniques and exiting use of DP in wearable data publishing. Based on our study we have identified the limitations of proposed solutions and provided future directions.

\end{abstract}
\end{@twocolumnfalse}
]

\section{Introduction}
Recent advances in wearable and smart technology, and the rapid adoption of wearable devices and smartphones makes them an important source of information for healthcare and medical research. The availability of these devices and the types of parameters they can measure is rapidly increasing. Real-time participant-generated physiological data can enable large scale observational studies of health conditions, provide better insights into the medical conditions of individuals, and help streamline clinical trial processes in medical research.

The researchers at IBM Watson stated than an average person possibly generates more than one million gigabytes of health-related data across his or her lifetime \cite{ibm}. These data are mostly physiological and personally identifiable data such as heart rate, blood pressure, respiratory rate, disease symptoms etc. These data generated from smart wearable healthcare devices have become a blessing for the modern healthcare. Whether remotely monitoring patients  or keeping track of physical condition and fitness, these data have the potential to transform the healthcare sector. The main difference between traditional healthcare data and the wearable device generated data is that data from wearables are of continuous nature which are updated dynamically and often can be temporally correlated. These data are used by medical professionals for continuous monitoring, researchers, analysts, insurance companies or sometimes even in health surveys \cite{paper420_7}. 

However, data privacy is a major concern for health data  \cite{paper420_282}. The need to ensure privacy and trust when sharing an individual's health data is particularly critical given the sensitive nature of health data and its protection by legislation (e.g. \cite{assistance2003summary} \cite{o2011regulation}). Breaches of such privacy are not uncommon \cite{grubb_2019}. The level of trust between participants in an open health data marketplace will be lower than, say, a hospital's clinical practice where the doctors and patients are known to each other. Therefore, appropriate mechanisms need to be established to ensure data security and integrity, and to build trust among the participants.

\begin{figure*}[t]
    \centering
    \includegraphics[width=0.85\paperwidth]{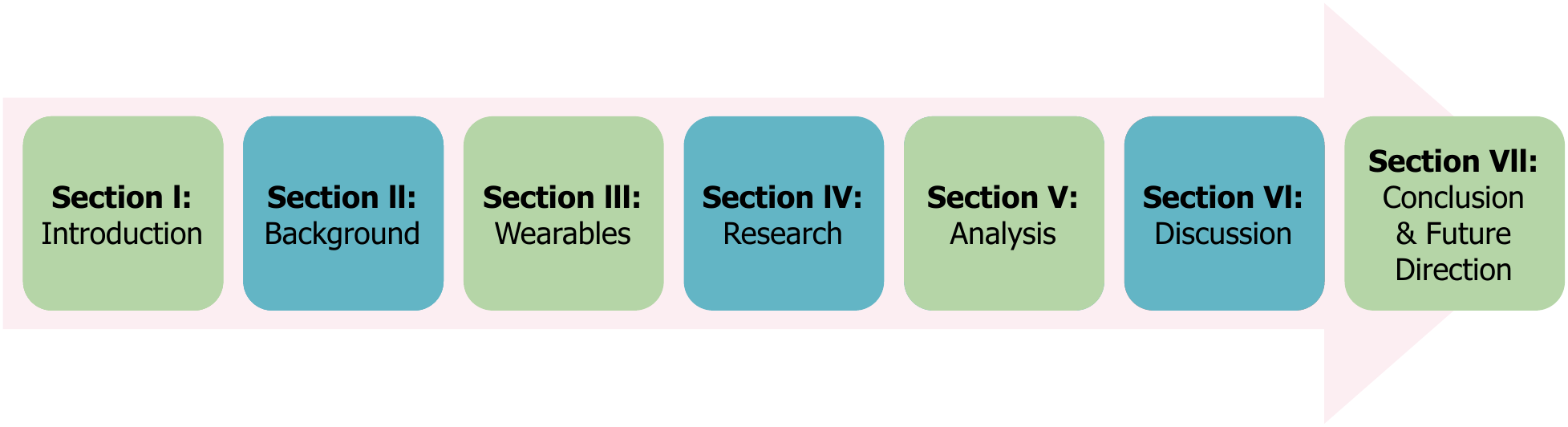}
    \caption{Organization of the paper}
    \label{fig_organization}
\end{figure*}

For preserving the privacy of sensitive data, many solutions have been proposed such as cryptography  \cite{paper_cryp1, chowdhury2009new}, blockchain  \cite{paper_bc1, paper_bc2, paper_bc3, paper_bc4, paper_bc6}, anonymization  \cite{paper_anony1, paper_anony2, paper_anony3, paper_anony4, paper_anony5}, privacy policy \cite{chowdhury2018policy}, and access control \cite{chowdhury2019continuous, chowdhury2018system}. All of these techniques and mechanisms have limitations, specially for publishing real-time dynamic data. A recent privacy technique, called Differential Privacy by Dwork  \cite{dwork_survey} has revolutionized researches in the privacy domain. DP ensures the privacy of the individuals in a way that the presence or absence of any individual in the published dataset cannot be discovered. This reduces the risk of privacy leakage of sensitive real-time data to a great extent \cite{soria2014enhancing}.  Therefore, it is used by different technology giants like Facebook, Google and Uber to protect the privacy of their user \cite{facebook_use_dp, google_use_dp, uber_use_dp}. 

One of the biggest challenges in real-time data is the high dimensional temporal correlation between data. Many researchers have found differential privacy suitable for preserving privacy in real-time health data and claimed that these solutions have advantages over existing methods. This proves that differential privacy is a fruitful mechanism and provides a more practical way for preserving privacy of real-time health data. To the best of our knowledge, there is no dedicated survey, traditional literature review, or a Systematic Literature Review (SLR) for privacy-preserving wearable physiological data publishing using DP. This motivates our work in this paper. 

We have performed an SLR on wearable data publishing (which generates data in a continuous manner) under differential privacy in the period from \textit{2007} to \textit{April 31, 2020}. We have come up with a holistic view of preserving privacy of wearable physiological data according to the existing literature. By performing a systematic mapping, we have analyzed the techniques, their use cases, datasets, experiment scenario and limitations of the existing solutions. We have categorized the research papers mainly into three major parts: \textit{Physiological, Real-time, and Others}. We have explored and analyzed how the research community have addressed them, how they have contributed by approaching different types of techniques, what experimental procedure they have considered and what limitations they have summed up.

We have illustrated the structure of this paper in Fig. \ref{fig_organization}.
In Section \ref{sec:background}, we have narrated necessary mathematical concepts of differential privacy, and their basic mechanisms. In Section \ref{sec:wearable}, we have discussed about wearable devices, types of data they generate and difference between wearable health data and traditional health data. We have explained the systematic literature review process in Section \ref{sec:research} and provided analysis in Section \ref{sec:analysis}. Section \ref{sec:discussion} has provided a brief discussion. Finally, we have concluded with future directions in Section \ref{conclusion}.

\section{Background} 
5

\label{sec:background}
In this section, we have provided the definition of differential privacy and its different variants. We have also discussed about different relevant concepts related to differential privacy. 
\subsection{Differential Privacy}
\begin{figure}
    \centering
    \includegraphics[width=0.41\paperwidth]{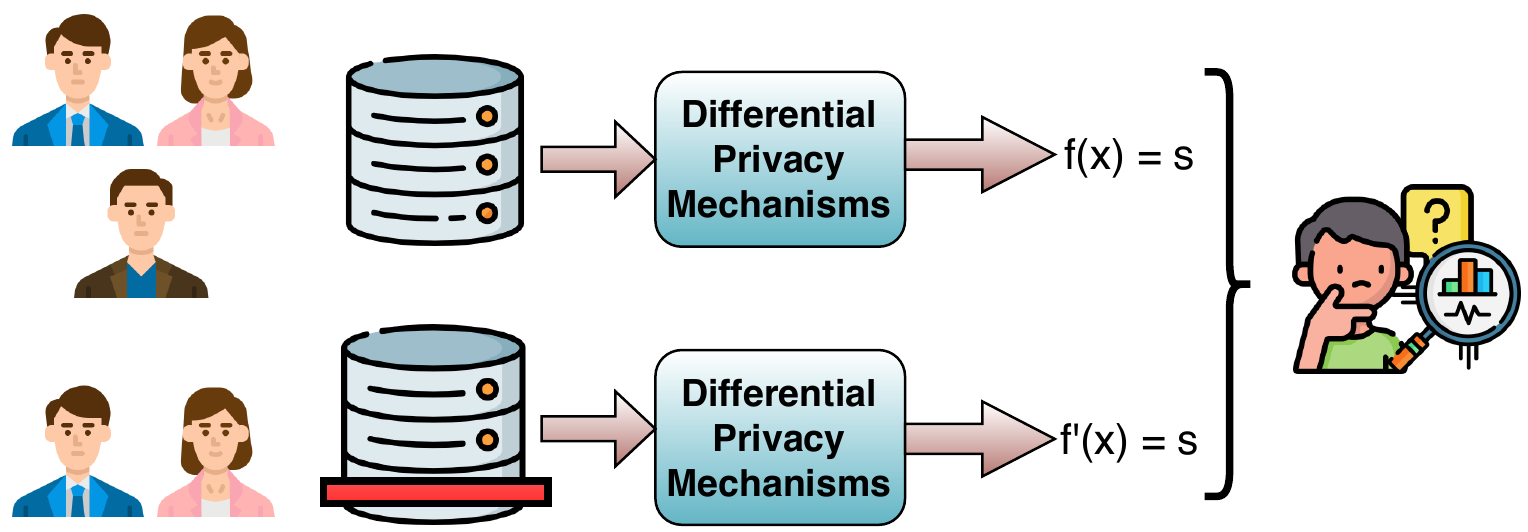}
    \caption{Differential privacy}
    \label{fig_dp}
\end{figure}

Differential privacy is the process of providing privacy of database in such a way that it should not reveal any Personal Identifiable Information (PII) about any individual for any query. In other words, nobody can ascertain the participation or non-participation of any individual in any dataset. The final result will not be affected by the presence or absence of any individual. Fig. \ref{fig_dp} shows a simple visualization of differential privacy. In the upper database, Munshi, Tajkia, and Bob have contributed their respiratory data. Eve (adversary) wants to find out Bob's respiratory rate. If we delete Bob's respiratory data, the probability of finding Bob's data in the database before and after deletion will be identical. Which means that, Eve can not ascertain whether or not Bob is included in the dataset, let alone the contents of his data. Hence, Bob's privacy is preserved.


\subsubsection{Definition}
A randomized mechanism $M$ gives $(\epsilon,\delta)$-DP for every set of outputs $S$, and for any neighbouring datasets (datasets that differ in only one value) of $D$, $D^{\prime}$ if $M$ satisfies Eq. \ref{eq1} \cite{dp9}:
\begin{equation}
\label{eq1}
\frac{\operatorname{Pr}[M(D) \in S]}{\operatorname{Pr}\left[M\left(D^{\prime}\right) \in S\right]} \leq e^{\epsilon}+\frac{\delta}{\operatorname{Pr}\left[M\left(D^{\prime}\right) \in S\right]}
\end{equation} 
This is known as \textit{approximate differential privacy}. If $\delta = 0$, then Eq. (\ref{eq1}) shows the ratio between the  probability of the output being into dataset $D$ and $D^{\prime}$ becomes less than or equal to $e^{\epsilon}$. This is known as \textit{pure differential privacy}. If two datasets 
differ with $c$ values then the ratio becomes less than or equal to $e^{\epsilon c}$.  This is known as \textit{group privacy}.

The mechanism $\epsilon$ and $M$ are the main actors here. $\epsilon$ is the balance between privacy loss and maximizing utility.
    \begin{enumerate}
        \item $\epsilon = 0$ leads to complete privacy but lack of utility.
        
        \item $\epsilon <= 1$ leads to less privacy but higher utility.
    \end{enumerate}
$M$ decides how much noise (i.e. a calibrated value used to anonymize data) will be added and what type of query is being served.

\subsubsection{Illustration of Differential Privacy}
\begin{figure}[!h]
    \includegraphics[width=0.41\textwidth, center]{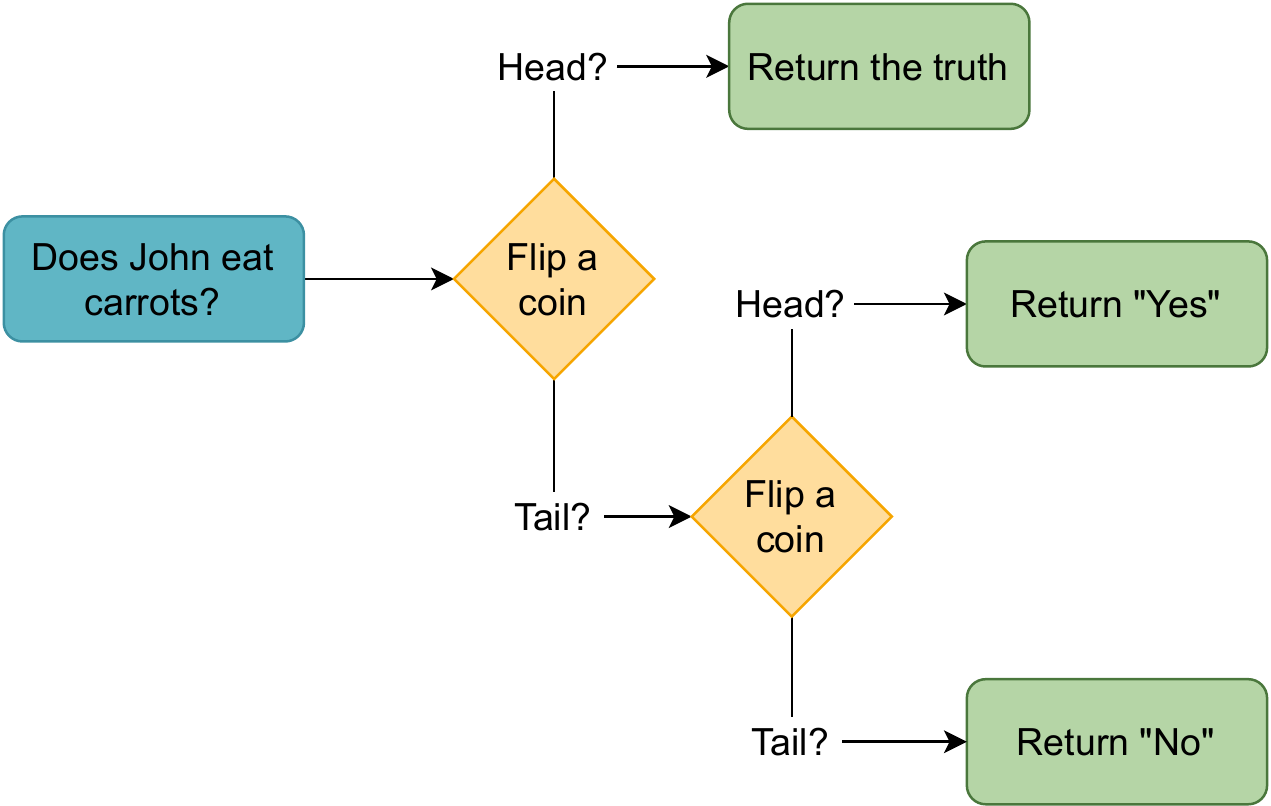}
    \caption{Flow diagram of the Differential privacy algorithm}
    \label{fig_dp_illustrate}
\end{figure}

Let us consider a child named John. There is a large amount of data of John which has been generated during John’s lifetime. This data is managed by John’s  parents. One example of such data is John’s preference: which veggies John dislikes? However, he is a bit ashamed about his preference and hence, may not wish for everyone to know this. The ability to keep this type of secret is referred to as privacy. However, the man who prepares John’s lunch at his daycare may wish to know that some of the children in the group dislike carrots! He is not required to know whether or not John is one of these children. It is sufficient if he is aware that there are perhaps four or five children who dislike carrots. This is referred to as differential privacy. 
Now, if the cook asks John whether he likes carrots or not, instead of giving a direct answer, he could utilize an approach which simulates differential privacy. In this approach, John will provide an indirect answer. More specifically, he will answer it through an outcome of some random process, e.g. a coin flip. The cook cannot see the result of this random process, this is essential. John flips the coin. The process is shown Fig. \ref{fig_dp_illustrate}. 

\begin{enumerate}
    \item If it is head, John will provide the true answer.
    \item If it is tail, John will flip the coin again.
    \begin{itemize}
        \item If it is head, John will say yes, regardless of what the true answer is.
        \item If it is tail, John will say no, regardless of what the true answer is.
    \end{itemize}
\end{enumerate}


This is an example of plausible deniability which refers to an individual's ability to deny anything since there is no concrete proof to show him right or wrong. Continuing with the example, the randomness of flipping a coin allows John to be protected with plausible deniability  as it is plausible for him to deny the answer based on the outcome of coin flipping. This randomized response process is actually differentially private process. Although the algorithms for differential privacy are much more complex, the principle remains the same. By making it unclear whether or not each response is legitimate, or even by altering replies arbitrarily, these algorithms can assure that regardless of how many queries are sent to the database, no one can be identified concretely.

\subsubsection{Local Differential Privacy (LDP)}
\begin{figure*}[t]
\includegraphics[width=0.8\textwidth, center]{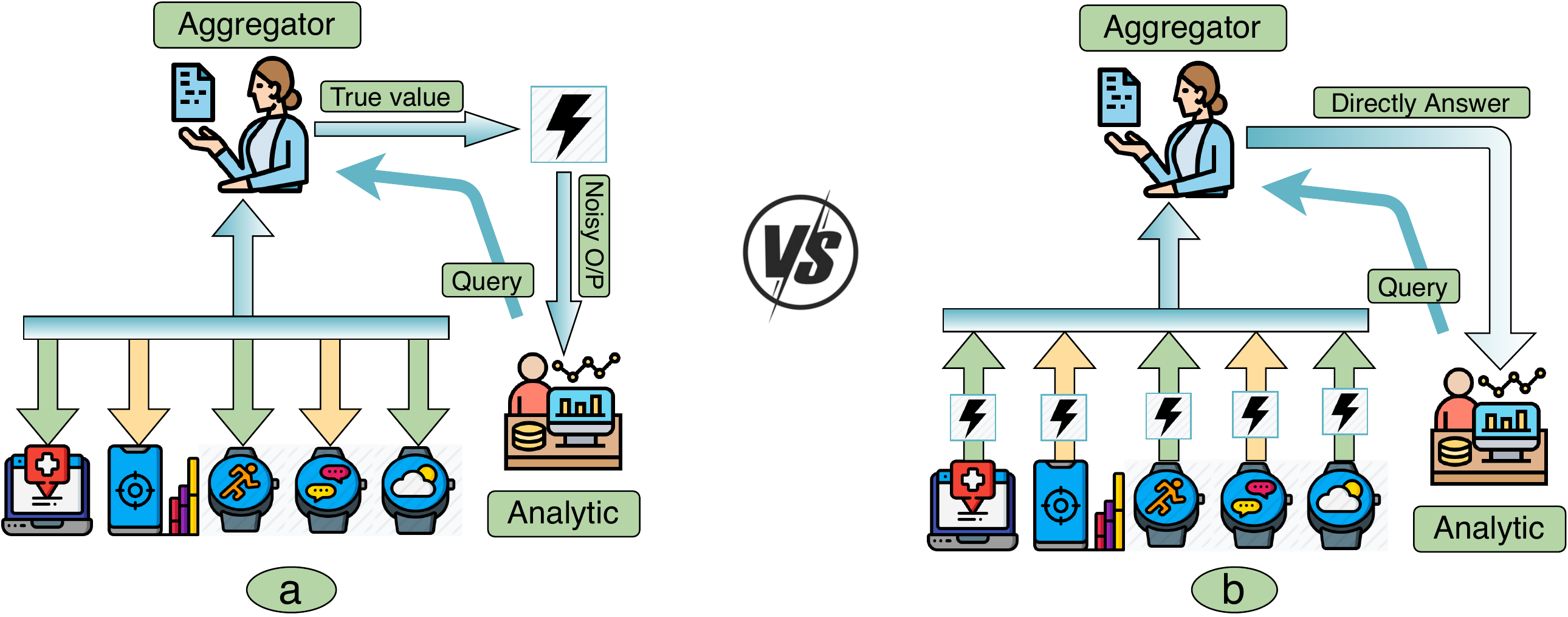}
\caption{Differences between (a) DP and (b) LDP}
\label{fig_dp_vs_ldp}
\end{figure*}

LDP \cite{dp2, dp3} is a slightly different method to achieve differential privacy. In other words, it is used to provide differential privacy locally. Differential privacy was designed for the purpose of sharing data, whereas LDP protects  the process of data collection  by maintaining individual privacy. In this case, a user instead of giving true data directly to the aggregator (an entity that collects as well as aggregates data from different sources and anonymize them), they add noise to their individual data first, then send the noisy data to the aggregator. This ensures individual privacy.  Fig. \ref{fig_dp_vs_ldp} shows the differences between DP and LDP. 

\subsubsection{Sensitivity}
Sensitivity is the maximum difference on query results between neighbouring datasets and defined as $\Delta f$ \cite{dp7}. Suppose there is a query "How many people in the database have the property $P$?". In this scenario, the presence or absence of an individual will change the result to a maximum value of only 1. So the sensitivity of this dataset is just 1. There are two types of sensitivity in differential privacy, namely \textit{Local} sensitivity and \textit{Global} sensitivity.

\subsubsection{Privacy Budget}
\label{subsubsec:privBudget}
\begin{figure}[h]
    \includegraphics[width=0.46\textwidth, center]{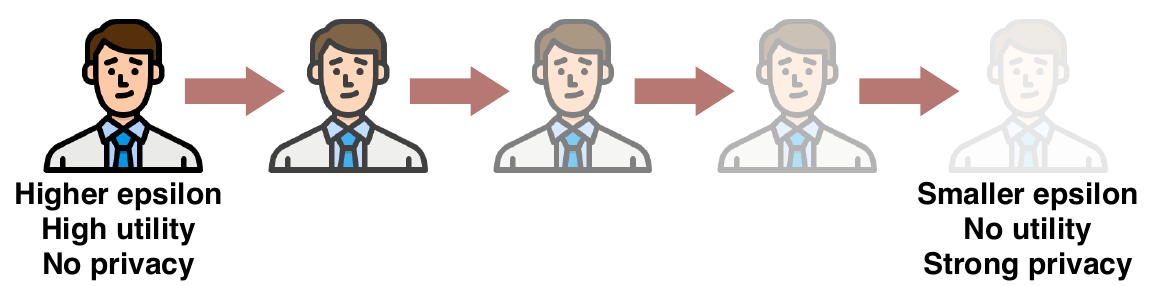}
    \caption{Effect of privacy budget}
    \label{fig_utility}
\end{figure}

$\epsilon$ is known as the privacy budget which controls the privacy guarantee level of any mechanism $M$ \cite{dp9}. The main responsibility of the privacy budget is to maintain the balance between privacy loss and utility maximization. Smaller $\epsilon$ (i.e more noise) ensures stronger privacy. But due to smaller epsilon the data can lose its utility and vice-versa. So, it is important to find and maintain the balance between privacy loss and utility maximization. Fig. \ref{fig_utility} shows a visual representation of this problem. As shown in figure, the more noise is added to the face image, the more anonymous it gets. But at the same time, with more anonymization the image becomes less useful. Similarly, less noise preserves utility of the image but does not provide any considerable privacy.



\subsection{Mechanisms of Differential Privacy}
\label{subsec:mechanismdiffprivacy}
Differential privacy can be achieved in two different ways, namely interactive and non-interactive. In interactive way, the system response to each query individually until total privacy budget is consumed. Whereas in non-interactive way, the data curator (who maintains and manages metadata) either evaluates and brings out statistics or discloses raw data anonymously. All the query responses are given at a time. In addition, DP uses different types of mathematical and statistical model for data perturbation based on the types of data such as numeric data and non-numeric data \cite{dp9}.

\begin{enumerate}
    \item For \textbf{numeric queries}, Laplace and Gaussian mechanisms are more suitable.
    \item For \textbf{non-numeric queries}, Exponential mechanism is more suitable.
\end{enumerate}

\subsubsection{Laplace Mechanism}
The \textit{laplace mechanism} is the procedure of adding \textit{laplace noise} to the query result \cite{dp7}. The noise is sampled from the \textit{laplace distribution} \cite{ dp9}. Eq. \ref{eq3} shows the probability density function for \textit{laplace distribution} which is centered at 0 with scale b:
 
        \begin{equation}
            \label{eq3}
            \operatorname{Lap}(x | b)=\frac{1}{2 b} \exp \left(-\frac{|x|}{b}\right)
        \end{equation}

The \textit{laplace mechanism} uses $l_1$-sensitivity (magnitude by which a single individual's data can change) and the variance of this distribution is $\sigma^{2}=2 b^{2}$. 

\subsubsection{Gaussian Mechanism}
In \textit{gaussian mechanism}, gaussian noise is added to the function \cite{dp9}. Rather than scaling to $\ell_{1}$-sensitivity, curator scales it to the $\ell_{2}$-sensitivity. Eq. \ref{eq5} shows the mechanism of adding \textit{gaussian noise} to the results.
        \begin{equation}
        \label{eq5}
            M(D)=f(D)+N\left(0, \sigma^{2}\right)
        \end{equation}

Where $\sigma=\Delta_{2} f \sqrt{2 \ln (2 / \delta)} / \epsilon$. And $N\left(0, \sigma^{2}\right)$ is the added \textit{Gaussian noise}.

\subsubsection{Exponential Mechanism}
The \textit{exponential mechanism} \cite{dp9} is used in case of non-numeric attributes because both the \textit{laplace mechanism} and \textit{gaussian mechanism} cannot deal with non-numeric attributes. In this case, the quality of an outcome is measured using a \textit{score function}. The score function $q(D,\phi)$ represents how good an output $\phi$ is for the dataset $D$. Eq. \ref{eq6} represents the equation of exponential mechanism. 

\begin{equation}
\label{eq6}
\begin{array}{l}
M(D)=\{\text { return } \phi \text { with the }
\text { probability \}} \\ \quad \quad \quad \quad \quad \quad \quad \quad \quad \quad \quad \quad \propto \exp \left(\frac{\epsilon q(D, \phi)}{2 \Delta q}\right)
\end{array}
\end{equation}

Where $\Delta_q$ represents the sensitivity of score function $q$.

\section{Wearables}
\label{sec:wearable}
    
Wearable devices can collect real-time data such as spatio-temporal data, trajectory data, location data but most importantly physiological data. These can track activities and remotely monitor a patient's condition. Consumer grade wearable trackers can track fitness and biosensors can collect biological data. Example of other wearables are smart footwear which includes smart shoes, socks, insoles and gloves \cite{paper_wearable}, smart jewellery that includes smart ring, smart bracelet even smart band, smart eye wear, glucose monitoring device, blood pressure monitor, body mounted sensor and biosensor.

\subsection{Wearable Device Architecture}

\begin{figure}[h]
 \centering
    \includegraphics[width=\linewidth]{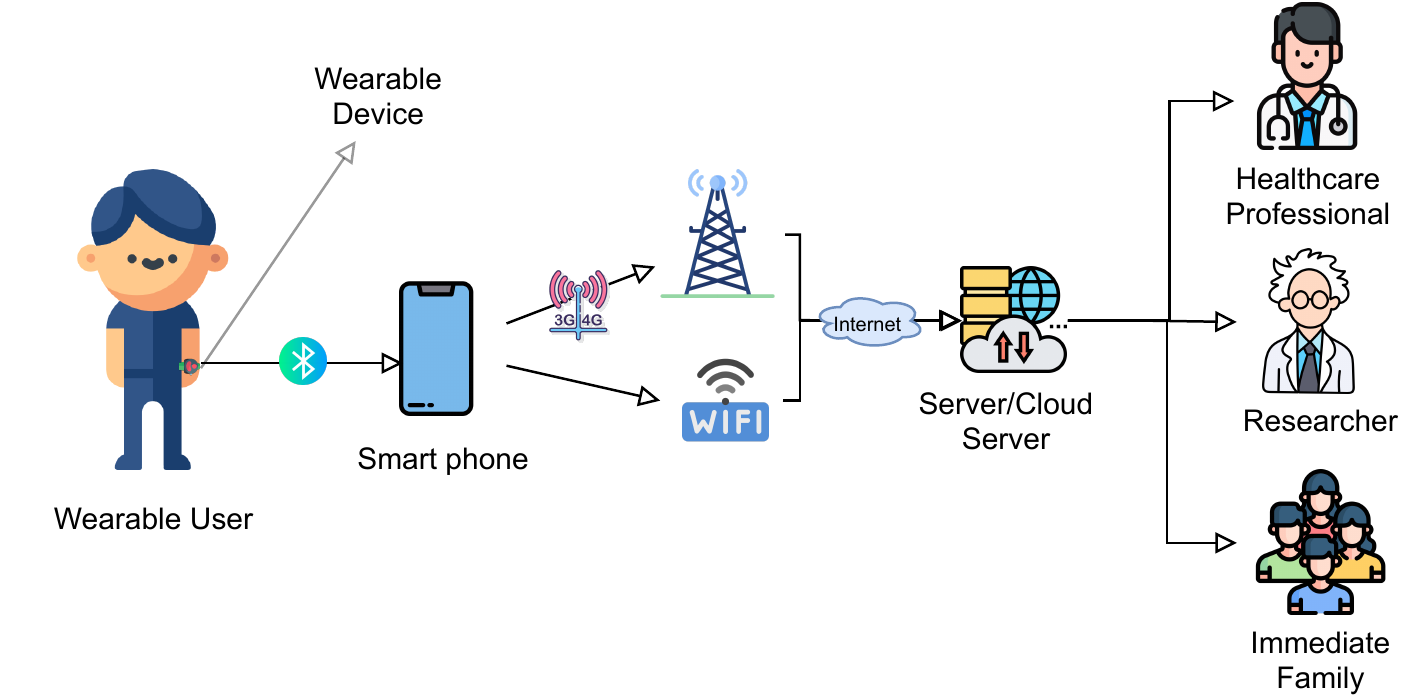}
    \caption{Architecture of a wearable device}
    \label{wearable_architecture}
\end{figure}

The architecture of wearable devices’ data exchange consists of three major components: 1) Wearable device 2) Smartphone and 3) Server/cloud server. Fig. \ref{wearable_architecture} illustrates the wearable scenario, demonstrating the data generation and sharing process. 

The wearable device collects data using sensors attached to it and transmits that data to the user’s smartphone. Bluetooth is used to transmit data between the wearable device and the smartphone. These sensors data are transmitted to the smartphone continuously and in real time. User applications on the smartphone enable the user to monitor the data collected by the wearable. After that, the data stored in the smartphone are transmitted to the remote server/cloud server via mobile network or WiFi. These server-stored data are shared with healthcare providers, health researchers, or immediate family members based on the user’s preferences.

\subsection{Types of Physiological Data}
Wearable device technology has become a crucial tool in the world of healthcare. These devices collect real-time physiological data continuously. Individuals can track their health condition and physical activities. Doctors can also monitor their patients remotely without having the patient to visit him/her in person or they can take necessary steps in case of emergency.

Some of the wearable devices are medical grade and some are customer grade devices such as Apple watch or Fitbit \cite{ref_different_grade_devices}. In recent times, the customer grade devices are becoming so sophisticated that they are now used in clinical trails or medical research. These devices are made of sensors and they use this sensing capability to identify different activities or physiological information. Several types of physiological data is collected by wearable devices. Some of them are listed below.

\begin{enumerate}
    \item \textbf{Heart rate:}
    Wearable devices like \textit{smart watch}, \textit{fitness trackers}, \textit{ECG monitors} and \textit{body mounted sensors} collect heart rate continuously. Some of the devices send notifications in case of any unusual response in heart rate.
    \item \textbf{Respiratory rate:} 
    \textit{Smart eye wear} and \textit{remote monitoring sensors} collect users respiratory rate.
    \item \textbf{Activity:}
    There are different wearable devices that track a user's movements and activities, calculate active minutes and sedentary minutes. Devices like \textit{smart band} (usually tracks fitness), \textit{smart health watch} and \textit{biosensors} collect physical activity and movement data.
    \item \textbf{Glucose level:} A \textit{glucose monitoring system} continuously collects the glucose level and notifies the user in case of any unwanted situation.
    \item \textbf{Steps taken:}
    Different devices work as personalized systems and count the step taken by users and help them in maintaining their health.
    \item \textbf{Blood pressure:}
    Blood pressure is collected by devices like \textit{fitness trackers} and \textit{remote monitoring systems}.
    \item \textbf{Stress level:}
    Stress level data collected by different type of wearable devices such as \textit{smart jewellery}.
    \item \textbf{Distance travelled:}
    Devices like \textit{fitness trackers} and \textit{smart health watches} collect the data of distance and elevation.
    \item \textbf{Calories burned:}
    \textit{Smart bands} and \textit{smart watches} collect the data of how much calories are burned and help the user in keeping track of their daily activities.
    
\end{enumerate}

\begin{figure}[t]
    \includegraphics[width=0.5\textwidth, center]{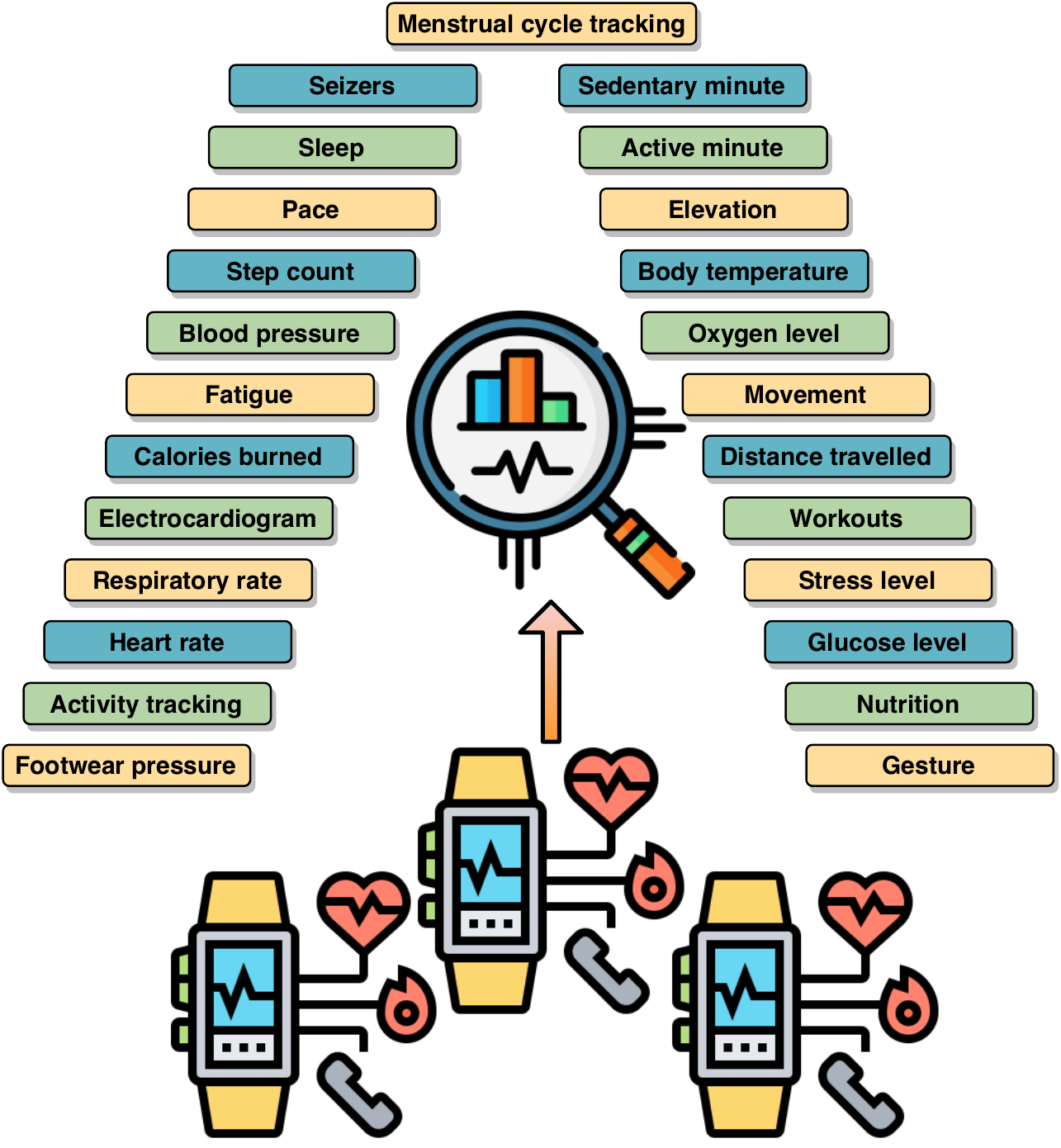}
    \caption{Physiological data collected by wearable devices}
    \label{fig_physiological_data}
\end{figure}

In addition to the above mentioned data, wearable devices also collect other physiological data such as oxygen level, menstrual cycle timing, body temperature and many more. Fig. \ref{fig_physiological_data} represents types of the physiological data generated and collected by different wearable devices.

\subsection{Difference between Traditional Health Data and Wearable Health Data}
\label{subsec:diffwearabledata}

Wearable data possesses some characteristics that differentiate them from traditional health data.
\begin{itemize}
    \item \textbf{Continuous data:} Wearable devices continuously generate data. Different sensors continuously capture different physiological data. The devices capture data even when we are sleeping. On the other hand, traditional health data are generated mainly by the health professionals when we visit them.
    \item \textbf{Numerical Data:} generated by the wearable devices are mainly numerical, such as blood pressure measurement, heart rate and number of steps walked. On the contrary, traditional health record keeps record of our health condition as diagnostic result or doctor’s interpretation of the diagnostics.
    \item \textbf{Time-series data:} Wearable devices data are stored as the time series data where as traditional health data are stored as textual format in the database.
    \item \textbf{Real time data:} Wearable devices provide the opportunity to capture the physiological data in real time. On the other hand, traditional data are static and are not real-time.
    \item \textbf{Highly correlated:} Data points in wearable data are highly correlated. 
    \item \textbf{More suitable for data analytic:} Wearable devices collect real-time data such as spatio-temporal data, trajectory data, location data but most importantly physiological data continuously (24/7) and in a format that is more consumable by the machine learning algorithms and can produce more accurate and effective data analytic.
\end{itemize}

\section{Research} 
\label{sec:research}

Through the SLR, we  have investigated the existing research which have attempted to apply differential privacy in wearable data and tried to overcome the challenges identified in the previous section. A SLR is methodologically rigorous in contrast to ad-hoc reviews \cite{blockchainSLR}. Our main focus is to identify relevant  papers and   review applications of differential privacy in wearable device generated physiological data, as well as to understand the conditions important for applying DP.

\subsection{Research Questions}
We have created five Research Questions (RQs), showed in Table \ref{tab:research_question} to guide our review. 

\begin{table}[ht]
\centering
\caption{Research questions}
\label{tab:research_question}
\begin{tabular}{ll}
\hline
\rowcolor[gray]{.9} \textbf{ID} & \textbf{Research Questions} \\ \hline \hline
 RQ1 & \begin{tabular}[c]{@{}l@{}}What are the DP techniques that have been used \\ in wearable data publishing? \end{tabular}  \vspace{0.17cm}       
 \\ 
\rowcolor[gray]{.96} RQ2 & \begin{tabular}[c]{@{}l@{}}What are the major contributions of the \\proposed solutions in wearable data publishing?\end{tabular} \vspace{0.17cm} 
\\
RQ3 & \begin{tabular}[c]{@{}l@{}}What types of datasets and programming languages \\ are being considered for evaluation and \\ implementation?\end{tabular}  \vspace{0.17cm}
 \\ 
\rowcolor[gray]{.96} RQ4 & \begin{tabular}[c]{@{}l@{}}What are the privacy criteria used in data \\publishing?\end{tabular} \vspace{0.17cm} 
\\ 
 RQ5 & \begin{tabular}[c]{@{}l@{}}What are the limitations of the proposed solutions?\end{tabular} \\ \bottomrule
\end{tabular}
\end{table}

\subsection{Search Strategy}
The overall search strategy is to find a body of relevant studies. Two search strategies, primary and secondary, have been used, as recommended  by  some  studies \cite{okoli2015guide, fink2019conducting} to  ensure that relevant studies have not been missed. In terms of record keeping, inclusion and exclusion strategies, we have followed PRISMA framework \cite{moher2011prisma} (detail numbers are in the Appendix \ref{ref:appendix}). For  the primary  search,  we have used  search  strings  on  several  electronic  databases.  Following  the  primary screening, we have conducted a secondary search (paper selection) by means of backwards and forward tracing. The primary screening strategy involves search terms, literature resources and search process. These are described briefly as follows. 

\subsubsection{Search Terms}
Throughout our searching process, we have considered journals and papers written in English. Besides this language factor, a date filter also has been  applied. 
We have conducted our searches in several digital libraries. We have maintained a conceptual research string containing the main keywords of the theme. The search keywords are given in Table \ref{tab:search_terms}. 
\begin{table}[h]
\centering
\caption{Search terms / keywords}
\label{tab:search_terms}
\begin{tabular}{ll}
\hline
\rowcolor[gray]{.9}\textbf{Number} &  \textbf{Keywords} \\ \hline \hline
1 & \begin{tabular}[c]{@{}l@{}}Review, survey, SLR, literature review\end{tabular}  
\\ 
\rowcolor[gray]{.96} 2 & \begin{tabular}[c]{@{}l@{}}Wearable, medical, health data\end{tabular} 
\\ 
3 & \begin{tabular}[c]{@{}l@{}}Wearable devices generated data\end{tabular} 
\\ 
\rowcolor[gray]{.96} 4 & \begin{tabular}[c]{@{}l@{}}Data publishing\end{tabular} 
\\ 
5 &  Privacy preserving 
\\ 
\rowcolor[gray]{.96} 6 & Differential privacy
\\
7 & Temporal data
\\
\hline
\end{tabular}
\end{table}

\subsubsection{Literature Sources}
In literature resources, we have conducted the searching process for papers on seven different electronic databases. During the paper collection process, we also have considered published journal names, published year, Computer Science Bibliographies, the title of the paper, the number of citations as well as the link of the paper.

After building conceptual search terms, we have used these keywords for finding journal papers, conference papers, and review papers in our considered electronic databases. Since different databases use different syntax for the search string, we have adjusted our search terms to accommodate with different databases. The search has been conducted on all the seven databases covering title, abstract, and keywords. 
The results of the search strings are given in Table \ref{tab:total_downloaded_papers}.
\begin{table}[h!]
\centering
\caption{Number of papers retrieved from each digital library}
\label{tab:total_downloaded_papers}
\begin{tabular}{ll}
\hline
\rowcolor[gray]{.9} \textbf{Digital Library} & \textbf{No of Returned Papers} \\ \hline \hline
Google Scholar & 14,435 \\
\rowcolor[gray]{.96}IEEE & 93 \\
Springer & 1,210 \\
\rowcolor[gray]{.96}ACM DL & 2,359 \\
ScienceDirect & 666 \\
\rowcolor[gray]{.96}JAMIA & 39 \\
PubMed & 17 \\ 
\hline \hline
\textbf{Total} & \textbf{18,819} \\ \bottomrule
\end{tabular}
\end{table}

\subsubsection{Search Process}
SLR needs to comprehensively search all relevant sources; therefore, we have defined the search process by dividing it into the following two phases. 
\begin{enumerate}
    \item \textbf{Initial Searching Phase:} Searched in the seven electronic databases separately, and then gathered the returned papers together with those from a set of candidate papers. With the given search strings in Table \ref{tab:search_terms}, we have used appropriate logical operators (i.e., 'AND' and 'OR') along with parenthesis and quotation mark to refine our search and hence retrieved all the papers (details are in appendix \ref{ref:appendix}).
    \item \textbf{Reference Searching Phase:} Scanned the reference lists of the relevant papers to find other relevant papers and then, if any, added them into the set.
\end{enumerate}

\begin{figure}[]
    \includegraphics[width=\linewidth, center]{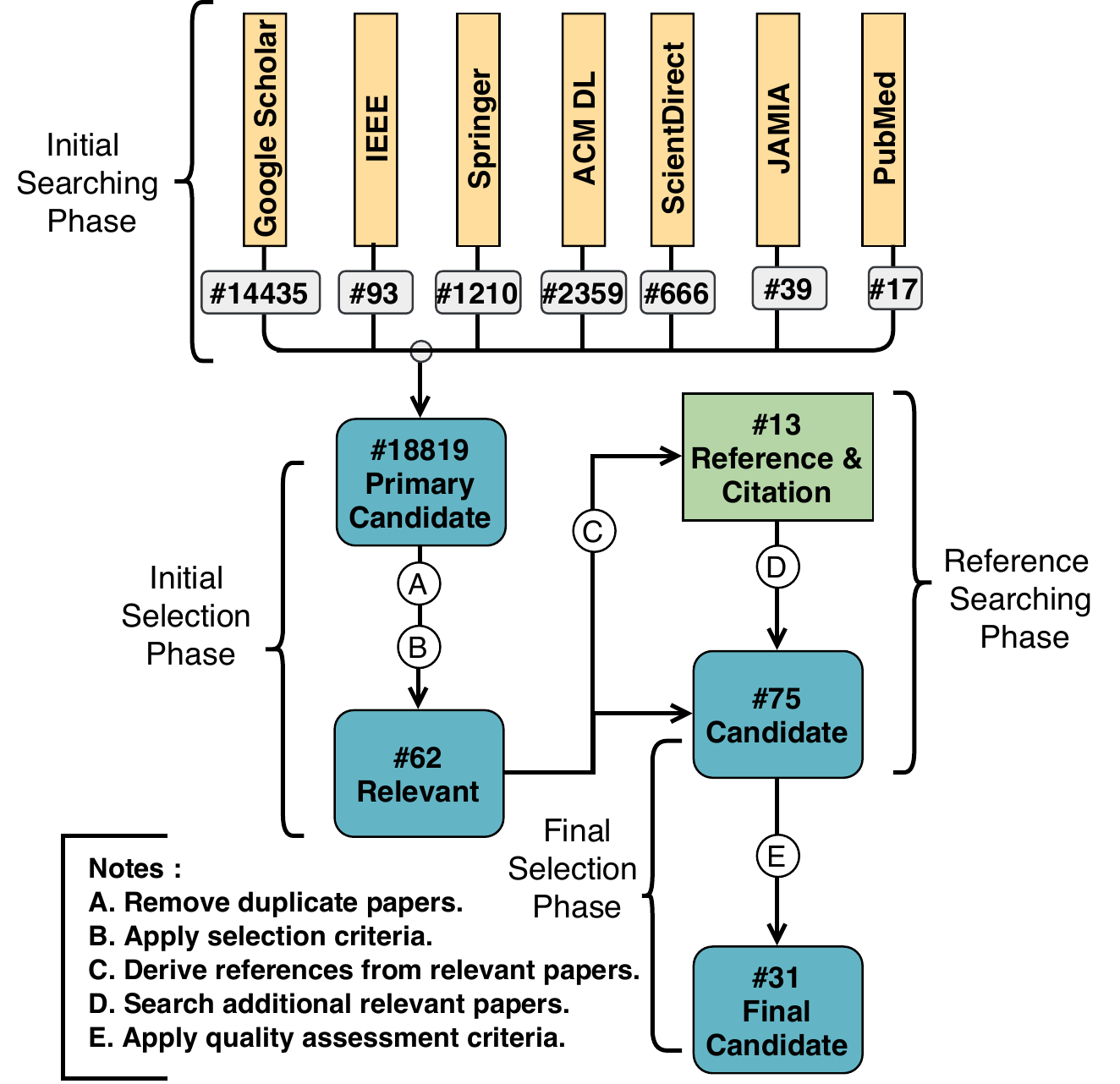}
    \caption{Search and selection process}
    \label{fig_search_process}
\end{figure}

We have used Microsoft Excel to store and manage the search results. We have gathered 18,819 papers from our initial searching phase and 13 papers from reference searching phase. Fig. \ref{fig_search_process} shows the search process in details including the number of papers.

\subsection{Study Selection}
We have found 18,819 candidate papers through Initial Searching Phase (see Fig. \ref{fig_search_process}). Since many of our candidate papers will not provide useful information to address the research questions raised by this review, we have conducted further filtering to identify the relevant papers. More specifically, the study selection procedure has the following two phases:
\begin{enumerate}
    \item \textbf{Initial Selection Phase:} We have applied the inclusion and exclusion criteria (defined below) to the candidate papers for filtering the relevant papers. These relevant papers can provide potential data for answering the RQs. 
    
    \item \textbf{Final Selection Phase:} In this phase, we have applied the quality assessment criteria (defined in Section \ref{research:study_quality_assessment}) to the relevant papers for selecting the papers with acceptable quality, which are eventually used for data extraction.
\end{enumerate}

Our inclusion criteria are presented below:
\begin{itemize}
    \item Abstract of papers written in English.
    \item Papers publish from 2007 onward. 
    \item Papers publish until April 31, 2020.
    \item Academic papers published in conferences or journals.
    \item Papers that describe wearable or real-time data publishing other than trajectory or location data.
    \item For duplicate publications of the same study, only the most complete and newest one is included.
    \item For study that has both conference version and journal  version, only the journal version is included.
    \item Review or Survey papers on real-time data publishing.
    \item Articles, Books related to wearable data publishing under differential privacy.
\end{itemize}

Next, the exclusion criteria are presented:
\begin{itemize}
    \item Abstract of papers written in other languages.
    \item Duplicated papers found on the digital libraries.
    \item Papers worked on static or traditional health data.
    \item Editorials, prefaces, summaries, interviews, news, correspondences, discussions, comments, reader’s letters, and summaries of tutorials, workshops, panels, and poster sessions.
\end{itemize}

We have extracted 62 papers through our inclusion and exclusion criteria. With that we have scanned the citation and references of these relevant papers and found 13 additional relevant papers which have been missed in the initial search process. Therefore, we have been able to identify 75 relevant papers. In the last step, we have applied a few quality assessment criteria for identifying the final selected studies, which are then used for data extraction.

\subsection{Study Quality Assessment}
\label{research:study_quality_assessment}
We have created some Quality Assessment Questions (QAQs) in order to validate the quality of the papers. These questions are showed in Table \ref{tab:quality_assessment_questions}.

\begin{table}
\centering
\caption{Quality assessment questions}
\label{tab:quality_assessment_questions}
\begin{tabular}{ll}
\hline
\rowcolor[gray]{.9} { \textbf{ID}} &
  { \textbf{Quality Assessment Questions}} \\ \hline \hline
{ QAQ1} &
  { \begin{tabular}[c]{@{}l@{}}Is the paper related to  real-time or wearable  data \\ publishing under  differential  privacy?\end{tabular}} \vspace{0.09cm} \\ 
\rowcolor[gray]{.96} { QAQ2} &
  { \begin{tabular}[c]{@{}l@{}}Is the dataset used in the experiment real\\ dataset or synthetic dataset? \end{tabular}} \vspace{0.09cm} \\  
 

{ QAQ3} &
  { \begin{tabular}[c]{@{}l@{}}Is the validation of the proposed method \\ done using a dataset or not?
  \end{tabular}} \vspace{0.09cm} \\ 
\rowcolor[gray]{.96}

{ QAQ4} &
  { \begin{tabular}[c]{@{}l@{}}Does the study add value to a digital library of \\ the industry community?\end{tabular}} \vspace{0.09cm} \\

{ QAQ5} &
  { Are the limitations of study analyzed explicitly?} \vspace{0.09cm} \\ 
\rowcolor[gray]{.96}{ QAQ6} &
  { \begin{tabular}[c]{@{}l@{}}Is the proposed publishing method compared with \\ other existing methods?\end{tabular}}  \\ \hline
\end{tabular}
\end{table}


The questions presented in \ref{tab:quality_assessment_questions} are used for the quality assessment of the research papers that we have collected. QAQ1 evaluates if the paper we are evaluating relates to wearable data publishing or real-time publishing or not. We have observed that researchers have evaluated their proposed DP-method using synthetic data rather than real wearable data; QAQ2 assesses this. If the dataset is a real wearable dataset, we keep them in our list otherwise, we discord them. In addition, some research works either do not have any validation or validation without a dataset; QAQ3 assesses that. QAQ4 evaluates if the research works adds any advance to the existing knowledge in terms of academic or industry practice. QAQ5 evaluates if the researchers have analyzed their own limitations or not. Finally, in QAQ6, we have checked whether the research paper has done any benchmarking with the existing work or not.

We have selected each paper based on the total number of QAQs they satisfy. We have read all 75 papers and evaluated them according to the Quality Assessment Questions. We have put a paper into final selected studies if the paper satisfies at least half of the QAQs. Finally, we have selected 31 papers. All the collected papers are listed in Table \ref{tab:selected_paper_list}.

\begin{table}
\centering
\caption{List of selected papers}
\label{tab:selected_paper_list}
 \resizebox{\linewidth}{!}{
\begin{tabular}{ll}
\hline 
\rowcolor[gray]{.9}\textbf{Category} & \textbf{Selected Papers}                                                         \\ \hline \hline
 Physiological  &
  \begin{tabular}[c]{@{}l@{}}Lin et al.  \cite{dp015}, Lin et al.  \cite{dp016}, Mohammad et al.\\ \cite{dp017}, Prema et al.  \cite{dp018}, Zhang et al.  \cite{dp019}, Song \\et al.  \cite{dp026}, Guan et al.  \cite{dp033},  Kim et al.  \cite{dp058}, \\Kim et al.  \cite{dp059}, Lin et al.  \cite{dp060}, Hao et al.  \cite{dp140}, \\Kim et al.  \cite{dp419}, Zhang et al.  \cite{dp422}, Julian et al.  \\\cite{dp439}, Zhang et al.  \cite{dp444}, Nazir et al. \cite{dp482},  Bozkir\\ et al. \cite{dp521} Arijit et al. \cite{dp139}.\end{tabular} \\

\rowcolor[gray]{.96} \vspace{6pt}Real-time &
  \begin{tabular}[c]{@{}l@{}}Liyue et al.  \cite{dp020}, Wang et al.  \cite{dp040}, Rastogi et al. \\ \cite{dp160}, Shi et al.  \cite{dp253}, Yang et al.  \cite{dp437}, Wang et al.\\  \cite{dp666}, Gao et al.  \cite{dp688}, Fan et al.  \cite{dp707}, Kellaris\\ et al.  \cite{dp709}. \vspace{6pt} \end{tabular} \\

  Others           & 
\begin{tabular}[c]{@{}l@{}}Nguy{\^e}n et al.  \cite{dp427}, Yang et al.  \cite{dp442}, Thomas et al.\\  \cite{dp459},  Luo et al.  \cite{dp469}. \end{tabular}               
  
   \\ \bottomrule
\end{tabular}
}
\end{table}

\section{Analysis}
\label{sec:analysis}

Before jumping into analyzing the RQs, we have organized the papers into following three categories. All the papers in these categories have used differential privacy to preserve the privacy of the health data.
\begin{enumerate}
\label{sec:categories}
    \item \textbf{Physiological} represents papers that have covered \textit{wearable physiological data}. Physiological data include, but are not limited to, EEG, ECG, EMG, blood pressure, and activity. These papers have mainly discussed how to collect these types of data using wearables and publish them in a privacy preserving way.  
    
    \item \textbf{Real-time} represents papers that have discussed  \textit{real-time} and \textit{dynamic} health data. This category only includes those papers which have discussed about real-time data collection or publishing. In terms of types of data, it is wearable heath data. However, it is exclusively for real-time health data collection and publishing. So, if a paper discusses about historical physiological data then we have placed it in the previous (physiological) category, however, if it discusses real-time physiological data then we have placed it under this category. 
    
    \item \textbf{Others} represents papers that are not directly related to wearables, rather they are related to Medical Internet of Things (MIoT) and smart devices, such as mobile phone based healthcare.

\end{enumerate}

\subsection{RQ1: What are the DP techniques that have been used in wearable data publishing?}
 
Wearable data has some characteristics which make them different from  traditional health data, such as real time, dynamic, numerical and highly correlated data(details in \ref{subsec:diffwearabledata}). We have discussed different types of differential privacy techniques in \ref{subsec:mechanismdiffprivacy}. In this research question, we will explore which of these techniques and other DP techniques have been used by the researchers to protect the privacy of wearable data. Here, we have reviewed the proposed techniques and methods to publish wearable physiological data under differential privacy.


Applying DP is challenging for real-time or transaction data, because differential privacy was built for providing $(\epsilon,\delta)$ privacy guarantee for statistical data only. Different researchers have addressed these issues and thus proposed different techniques for publishing real-time data while maintaining differential privacy. In traditional differential privacy mechanisms, it is assumed that the data are independent, i.e. they are not correlated and the adversary does not have any knowledge of the data correlations. But real-time generated data can be correlated or we can acquire correlations among data.

\begin{figure}[b]
    \includegraphics[width=0.50\textwidth, center]{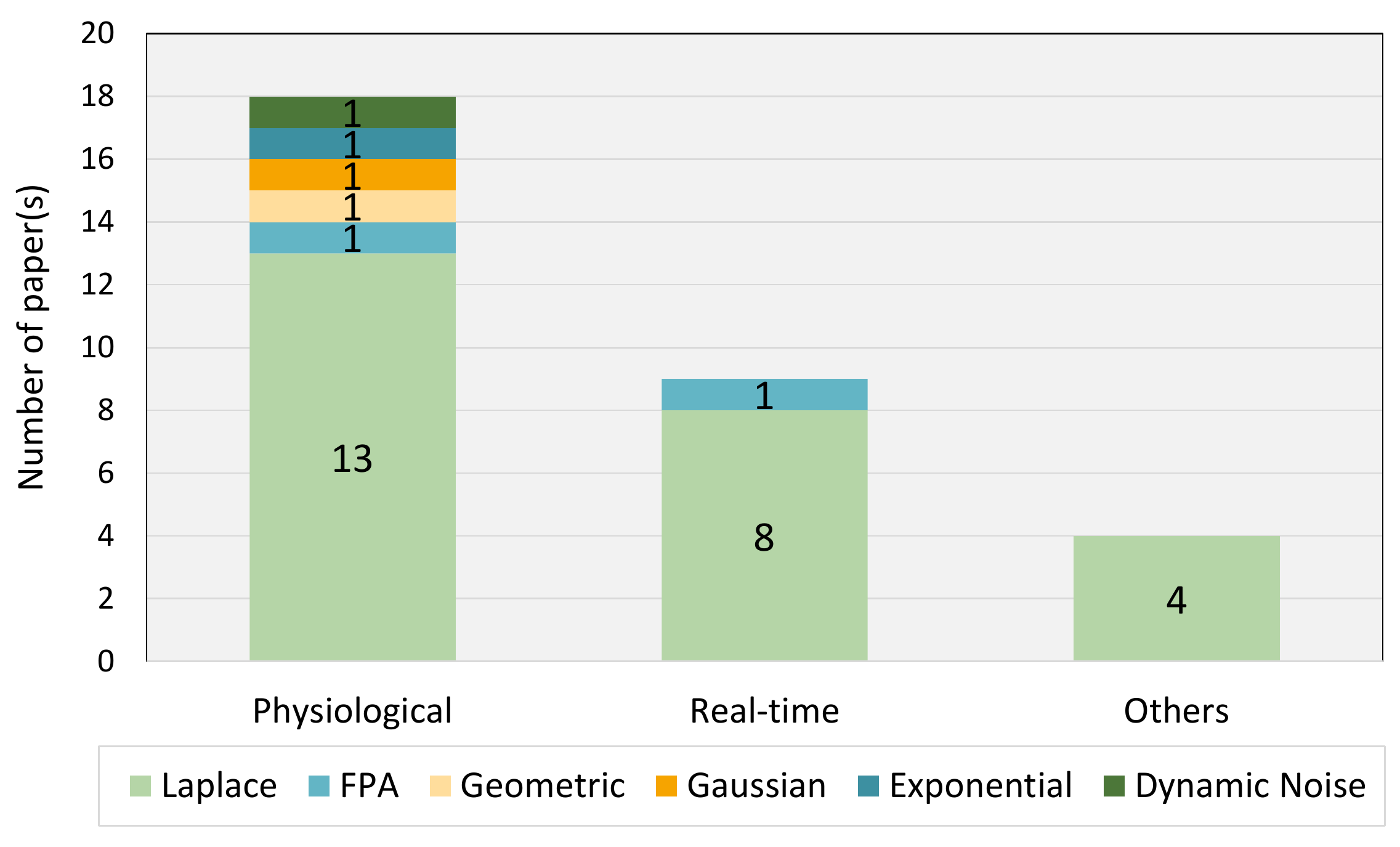}
    \caption{Considered perturbation techniques for achieving differential privacy}
    \label{fig_dp_techniques_bar}
\end{figure}

Fig. \ref{fig_dp_techniques_bar} represents different types of DP techniques used for different types of wearable data (e.g., physiological, real-time, and others). It is evident from the figure that among all the techniques Laplace Distribution is the most popular for adding noise to the data. Other techniques are Geometric Distribution and Fourier Perturbation Algorithm (FPA). For eye data, researchers have preferred Gaussian noise for perturbation  \cite{dp439, dp521}. For ensuring privacy guarantee of real-time and physiological data in health, techniques such as adaptive sampling  \cite{dp019, dp020, dp666, dp707}, filtering  \cite{dp019, dp020, dp666, dp707}, adaptive budget allocation  \cite{dp019, dp666}, filtering with Laplace distribution  \cite{dp666} have been used. 
 
\begin{table*}[t]
  \begin{center}
    \caption{Research work related to physiological data publishing using DP}
    \label{tab:rq2physio}
    \begin{tabular}{M{24mm} M{13mm} p{44mm} p{85mm}}
      \hline
      \rowcolor[gray]{.9} \textbf{Paper} & \textbf{Name} & \textbf{Used technique (For RQ1)} & \textbf{Major Contribution (For RQ2)}\\
      \hline \hline
       Lin et al.  \cite{dp015} & - & DP-based Dynamic Noise threshold & A DP-based new scheme for large data from body sensor networks\\\hline 
       \rowcolor[gray]{.96} Lin et al.  \cite{dp016} & - & DP (Laplace noise) with Haar Wavelet technique & Differentially private scheme for sensitive big data in BSNs with reduced errors. \\\hline 
       Mohammad  et al.  \cite{dp017} & - & Bucket algorithm and Laplace distribution & An efficient differentially private mechanism for releasing health data \\\hline 
       \rowcolor[gray]{.96} Prema et al.  \cite{dp018} & - & Selective Gaussian mechanism with DP & A DP-based scheme for big data BAN which is more available and reliable \\\hline 
       Zhang et al.  \cite{dp019} & RE-DPocpor & Laplace noise with adaptive sampling, filtering and budget allocation techniques & Dataset privacy where data that has been collected from $w$-consecutive days \\\hline 
       \rowcolor[gray]{.96} Song et al.  \cite{dp026} & PPM-HDA & Geometric Distribution & A more diverse and secure mechanism  resisting differential attacks and supporting additive and non-additive aggregation\\\hline
       Guan et al.  \cite{dp033} & EDPDCS & K-means clustering and Laplace noise based DP & Proposed an efficient privacy-preserving clustering scheme over Map-reduce Framework for IoMT \\\hline  
       \rowcolor[gray]{.96} Kim et al.  \cite{dp058} & - & Laplace distribution & Presented method is capable of preserving privacy of individuals sensitive data generated from smartwatches \\\hline 
       Kim et al.  \cite{dp059} & - & Laplace distribution by leveraging LDP & Developed method can collect sensitive health lifelogs from smartwatch under DP \\\hline 
       \rowcolor[gray]{.96} Lim et al.  \cite{dp060} & - & LDP with Laplace noise & Developed technology is capable of collecting health data from smartwatches by maintaining LDP \\\hline 
       Hao et al.  \cite{dp140} & PMHA-DP & Laplace noise with a hierarchical method & Proposed multi-functional health data aggregation scheme under DP \\\hline
       \rowcolor[gray]{.96} Kim et al.  \cite{dp419} & - & LDP with Laplace & Proposed mechanism can collect individual temporal health data at fixed intervals by leveraging LDP \\\hline 
       Zhang et al.  \cite{dp422} & WSV-MDAV & Laplace noise with micro aggregation algorithm & Proposed a privacy protection model based on aggregation algorithm for wearable devices using DP \\\hline 
       \rowcolor[gray]{.96} Julian et al.  \cite{dp439} & - & Exponential Mechanism & Designed interface for VR to prevent user re-identification and protect gender information using DP\\\hline 
       Zhang et al.  \cite{dp444} & APDP & Modified Laplace Mechanism & A fog computing based secured smart-home model with a personalized DP scheme\\\hline 
       Arijit et al.  \cite{dp139} & - & Data driven technique \& Laplacian noise for data obfuscation & On-demand obfuscation of sensitive data by satisfying DP \\
       \rowcolor[gray]{.96} Nazir et al.  \cite{dp482} & mSieve & Laplace Distribution & Defined a new behavioral privacy metric under differential privacy\\\hline  
       Bozkir et al.  \cite{dp521} & - & Fourier Perturbation Algorithm & A chunk based privacy-preserving method for eye movement features by considering different factors\\
       \hline 
    \end{tabular}
  \end{center}
\end{table*}

\begin{table*}[t]
  \begin{center}
    \caption{Research work related to real-time data publishing using DP}
    \label{tab:rq2Real}
    \begin{tabular}{M{26mm} M{12mm} p{48mm} p{78mm}}
      \hline
       \rowcolor[gray]{.9} \textbf{Paper} & \textbf{Name} & \textbf{Used technique (For RQ1)} & \textbf{Major Contribution (For RQ2)}\\
      \hline \hline
        Rastogi et  al.  \cite{dp160} & PASTE & Fourier Perturbation Algorithm (FPA) \& Distributed Laplace Perturbation Algorithm (DLPA) & Combining FPA and DLPA to achieve the accuracy benefits of the former and the scalability of the latter \\\hline
       \rowcolor[gray]{.96}
       Shi et al.  \cite{dp253} & PSA & Symmetric Geometric Distribution (SGD) with Laplace distribution & Combining differential privacy and cryptography to calculate the approximate aggregate statistics for a time interval over encrypted data \\\hline
       Liyue et al.  \cite{dp020} & FAST & Laplace noise (using a white Gaussian error with variance) & Improved data accuracy using Kalman filter and privacy cost minimization using adaptive sampling \\\hline
       \rowcolor[gray]{.96}
       Fan et al.  \cite{dp707} & FAST & Laplace noise with filtering and adaptive sampling & Differential private real-time aggregate statistics based on filtering and adaptive sampling \\\hline
        Kellaris et al.  \cite{dp709} & BA, BD & Laplace noise with sampling and dynamic privacy budget allocation & Proposal of two novel mechanisms along with several optimizations \\\hline
       \rowcolor[gray]{.96} Wang et al.  \cite{dp040} & UKFDP & Laplace noise with unscented Kalman Filter & Kalman filter based DP for nonlinear systems enabling differentially private streaming data share \\\hline
        Yang et al.  \cite{dp437} & ConTPL & Laplace mechanism & A system to automatically convert an existing differentially private streaming data within a specific level  \\\hline
       \rowcolor[gray]{.96}
       Wang et al.  \cite{dp666} & RescueDP & Laplace noise with adaptive sampling, budget allocation, dynamic grouping and filtering & Monitoring online aggregations of infinite streams with privacy guarantee \\\hline
       Gao et al.  \cite{dp688} & - & Laplace mechanism with GGA algorithm and Kullback Leibler (KL) divergence &  Proposed approach can publish histogram for differentially private dynamic data based on Kullback-Leibler (KL) divergence \\\hline
       
    \end{tabular}
  \end{center}
\end{table*}

Researchers have extended Laplace distribution to provide better privacy guarantee. Shi et al.  \cite{dp253} used Symmetric Geometric Distribution (SGD) with Laplace distribution to provide discrete approximation to the Laplace distribution. Haar Wavelet technique  \cite{dp016}, Bucket partition algorithm for partitioning dataset  \cite{dp017}, Geometric technique  \cite{dp026} have also been adopted by different researchers. Despite the temporal correlation, researchers such as Rastogi et al.  \cite{dp160} and Bozkir et al.  \cite{dp521} have tried to achieve differential privacy using FPA. 

In general, Laplace mechanism provides better accuracy compared to Gaussian mechanism. Therefore wearable data being mostly numerical, the Laplace mechanism may be an appropriate choice for perturbation.

\begin{table*}[hb]
  \begin{center}
    \caption{Research work related to others categories}
    \label{tab:rq2Others}
    \begin{tabular}{M{24mm} M{16mm} p{46mm} p{80mm}}
    \hline
     \rowcolor[gray]{.9} \textbf{Paper} & \textbf{Name} & \textbf{Used technique (For RQ1)} & \textbf{Major Contribution (For RQ2)}\\
      \hline \hline
       Nguy{\^e}n et al.  \cite{dp427}& Harmony & Local Differential Privacy (LDP) & An efficient solution for smart device data using LDP \\\hline
       \rowcolor[gray]{.96} Yang et al.  \cite{dp442} & MLDP & Machine learning with Laplace noise(noise added in training set) & A ML based differentially private aggregation method in IoT within a fog computing architecture to reduce communication overhead and release cloud burdens\\\hline
       
       Thomas et al. \cite{dp459} & - & Laplace distribution and Sine polyonym & Utility maximization with adjustable privacy settings for calculating aggregations over private sensor data\\\hline
       
       \rowcolor[gray]{.96} Luo et al.  \cite{dp469} & Salus, P3 & Dynamic Noise by leveraging Laplace Distribution & An input perturbation algorithm to preserve DP by providing strong resilience against data reconstruction attacks and predictable utilities \\ \hline
    \end{tabular}
  \end{center}
\end{table*}

\subsection{RQ2: What are the major contributions of the proposed solutions in wearable data publishing?}
In this section, we have explored the major contributions of DP-based techniques by different researchers in terms of reliability, utility, accuracy and risk minimization. We have divided the research works in three different categories.

\vspace{2mm}
\noindent \textbf{Physiological category:}

In  \cite{dp015}, authors have introduced a new DP-based concept called \textit{Dynamic noise threshold} which is suitable for large amount of data from Body Sensor network (BSN). Similarly, in  \cite{dp016}, authors have used differential privacy \\(Laplace noise) along with the \textit{Haar wavelet technique} \cite{haar_wavelet} for histogram to binary tree conversion to be used for sensitive big data in BSNs. Using this approach, they have been able to reduce errors and provide long-range queries using a tree structure. Prema et al.  \cite{dp018} have introduced another DP-based Gaussian scheme for BSNs in which Gaussian noise is applied to only important features if DP does not produce a satisfactory protection. They have claimed their approach is \textit{more reliable}. 


In terms of improving the \textit{utility}, Mohammad  et al. \cite{dp017} have proposed an efficient differentially private mechanism which adopts the bucket partition algorithm and Laplace distribution for preserving privacy. The effectiveness of their proposed technique is demonstrated by the overall improvement in the \textit{accuracy} of perturbed data. In a similar domain, Zhang et al.  \cite{dp019} have adopted Laplace noise as the data perturbation method along with adaptive sampling, filtering and budget allocation techniques. Their method allows the release of real-time health data with $w$-day differential privacy where the health data is collected for any consecutive $w$ days. They have compared their technique to state-of-the-art methods for performance comparison and have proved that their method outperforms others in terms of \textit{utility and privacy} guarantee.

To improve the \textit{accuracy} of DP-based techniques, the authors in  \cite{dp058, dp059, dp060} have adopted the Laplace distribution  noise as their base perturbation technique. According to \cite{dp058}, authors have asserted that their proposed approach can be used to \textit{efficiently} compute population data while maintaining privacy through the use of a wristwatch. However, the authors in  \cite{dp059, dp060} have additionally leveraged LDP for collecting health data from smartwatches. Both the research works have conclusively contributed to successfully preserving the \textit{usefulness} while properly gathering data from smartwatches with \textit{privacy preservation} in place. Finally, Kim et al.  \cite{dp419} have proposed a novel mechanism to collect individual temporal health data at fixed intervals by leveraging (by using max advantage) Laplace Differential Privacy. As a result, their proposed technique have outperformed straightforward methods by delivering a significant improvement in \textit{accuracy}.


Risk of privacy attack is major concern for any privacy preserving technique. Some researchers have worked to improve DP-based technique to \textit{avoid the attack}. For instance, an aggregation scheme named \textit{PPM-HDA} have been proposed in  \cite{dp026} which supports both multi-functional additive (average, variance) and non-additive aggregation (min/max, median, sigma-percentile, and histogram). It is claimed that the proposed mechanism is more diverse and secure for cloud servers, \textit{resisting differential attacks}. Authors in  \cite{dp033} have proposed a clustering scheme which introduces a privacy-preserving clustering scheme for the Map-reduce framework with\textit{ improved accuracy} by optimizing privacy budgets. To achieve this, they have used K-means clustering and Laplace noise for DP. Zhang et al.  \cite{dp444} have proposed a fog computing based smart-home model and explored collision attacks under personalized protection scenarios using DP. In their proposed model, noise is generated under a Markov process and the \textit{privacy protection} is achieved using a modified Laplace distribution. Their experiment have resulted in successful \textit{privacy enhancement} while minimizing overall privacy budget and \textit{eliminating background knowledge attack}. Authors in  \cite{dp422} have identified and solved the issues of V-MDAV algorithm and then, have proposed a privacy protection model, based on an aggregation algorithm, named \textit{WSV-MDAV} for wearable devices using DP. According to their experimental assessment, their technique have \textit{improved privacy protection performance} and \textit{reduced data loss} when compared to the traditional method. In  \cite{dp139}, Arijit et al. have proposed a solution that can obfuscate any sensitive data on-demand by satisfying differential privacy. This work is of practical importance that have \textit{improved the performance} by minimizing the privacy breaching risk.


\textit{Reduction of computational overhead} is also an active research direction. In  \cite{dp439}, authors have designed a Virtual Reality (VR) interface which \textit{prevents user re-identification} as well as \textit{protects gender information} by using DP. Their experiment is effective in \textit{reducing overhead}, resulting in a \textit{low-cost solution} for preserving users' privacy while preserving utility. In  \cite{dp521}, Bozkir et al. have put forward a chunk based privacy-preserving method for eye movement features by considering the factors reduction of query \textit{sensitivity}, \textit{complexity} and \textit{temporal correlations}. Their transform coding based solution is claimed to be \textit{more adaptive }than various existing low-complexity methods. Both these papers are related to eye movement data.

A summary of the used techniques and major contributions in the research papers under the physiological category is presented in Table \ref{tab:rq2physio}.

\vspace{2mm}
\noindent \textbf{Real-time category:}


In  \cite{dp160}, authors have proposed a scheme for real-time health data named PASTE where both the Fourier perturbation algorithm and Laplace distribution are used. By perturbing Discrete Fourier Transform (DFT) of query answers, the proposed FPA algorithm can answer multiple queries over time-series data and ensure DP despite the presence of a temporal correlation. On the other hand, the proposed DLPA (Distributed Laplace perturbation Algorithm) can be used for adding noise in a distributed way, a useful feature in the absence of a trusted third party. By combining FPA and DLPA, PASTE gets the\textit{ accuracy benefits} of the former and the \textit{scalability} of the latter. In  \cite{dp253}, authors have proposed a solution by combining differential privacy and cryptography enabling a user to upload a stream of encrypted data to an aggregator (can be untrusted) and the aggregator can calculate the approximate aggregate statistics for a time interval through the proposed algorithm. Combining these methods have helped them achieving \textit{strong privacy} guarantee. 


Some researchers have used adaptive sampling and filtering to preserve the privacy of real-time data to improve \textit{utility and performance}. For example, in  \cite{dp020}, the proposed approach enables to release time-series data under differential privacy by \textit{improving data accuracy} (using Kalman filter \cite{kalman_filter}) and \textit{minimizing overall privacy cost} (adaptive sampling algorithm with PID control). Similarly, Fan et al.  \cite{dp707} have proposed a framework to release real-time aggregate statistics by satisfying differential privacy based on filtering and adaptive sampling. Their adaptive methods \textit{improves the utility} and demonstrates excellent \textit{performance} even under small privacy cost. 

In  \cite{dp709}, authors have considered sliding window methodology using Laplace noise with sophisticated sampling and dynamic privacy budget allocation techniques. This will improve the \textit{scalability} in terms of real time data publishing. They have also proposed three benchmark methods named \textit{FAST\textsubscript{w}}, \textit{Uniform}, and \textit{Sample}. The solution is based on Kalman filter based differential privacy to facilitate streaming data sharing. It takes advantages of sigma points  \cite{dp040} for non-linear systems. This method \textit{increases the accuracy} of the published data \cite{dp020}. 


To \textit{overcome the temporal correlation} problem in real-time data, Yang et al. \cite{dp437} have designed system that can automatically convert an existing differentially private streaming data into one bounding Temporal Privacy Leakage (TPL). On demand sensitive data obfuscation is also used for real-time streaming data. Furthermore, authors in  \cite{dp666} have proposed a framework named RescueDP which can monitor the online aggregation of an infinite stream with privacy guarantee. Using adaptive and dynamic methods RescueDP outperforms existing methods and \textit{preserves utility} with proper \textit{privacy guarantee}.

Finally, the proposed algorithm by Gao et al.  \cite{dp688} can publish histograms for differentially private dynamic data based on Kullback-Leibler (KL) divergence \cite{kullback-leibler}. Using this methods have resulted in overall \textit{accuracy improvement} and \textit{utility enhancement}. 

A summary of the used techniques and major contributions in the research papers under the real-time category is presented in Table \ref{tab:rq2Real}.

\vspace{2mm}
\noindent \textbf{Others category:}
Thomas et al. \cite{dp459} have proposed to select the proper privacy settings to calculate an aggregation function over private sensor data. It can also help to \textit{maximize the utility} for different level of privacy. This makes the method \textit{secure and reliable}. In  \cite{dp442}, authors have considered a fog architecture instead of cloud architecture, where their proposed multi-functional aggregation method \textit{reduces communication overheads} as well as \textit{releases cloud burdens}. Considering data reconstruction attacks, Luo et al.  \cite{dp469} proposed an input perturbation algorithm \textit{Salus}. This light-weight algorithm provides \textit{strong resilience} against data reconstruction attacks while preserving differential privacy. Later \textit{Salus} was extended in P3 framework for supporting privacy-preserved Mobile Crowdsensing Services (MCS) applications \cite{dp442}. Authors in  \cite{dp427} have proposed a system that is \textit{practical, accurate, and efficient} for gathering and examine data from smart device users under LDP.


If we review the key contributions of researchers to differentially private health data publication discussed in this article, we can demonstrate that the most important aspect that have been considered by researchers are \textit{privacy and utility enhancement}. Due to the fact that differential privacy involves a trade-off between privacy and utility, it is critical to address privacy preservation in a way that does not jeopardize the utility of data. Additionally, researchers have concentrated on improving the \textit{overall performance} of their proposed mechanism in order to outperform existing works. Improving mechanism \textit{accuracy} have also been a center of focus of the researchers. Additionally, the researchers have focused on\textit{ overhead reduction} and \textit{secure and reliable} model building. Although It has been noticed that researchers have placed a greater emphasis on utility, privacy and performance enhancement compared to the model's security and reliability.

A summary of the used techniques and major contributions in the research papers under the physiological category is presented in Table \ref{tab:rq2Others}.

\subsection{RQ3: What types of datasets and programming languages are being considered for evaluation and implementation?}
Validating any research proposal is a very important part of the research. Researchers employ different types of methods to validate their proposed methods, systems, protocols, or techniques. In a data-driven scenario, the quality of validation heavily depends on the quality of the dataset. In this section, we have reviewed the types of dataset the researcher have considered to evaluate their research proposal. We have divided the dataset into two categories based on their availability, such as

\begin{enumerate}
    \item \textbf{Public datasets:} Datasets that are publicly available.
    
    \item \textbf{Private datasets:} Datasets that are self made or self collected by the researcher. We have considered synthetic datasets also a private datasets. 
\end{enumerate}

\vspace{2mm}
\noindent \textbf{Physiological category:} In the physiological category, a large variation of datasets have been utilised. In terms of heart related dataset, Zhang et al.  \cite{dp019} have collected heart rates data for three months from hospital patient. Nazir et al.  \cite{dp482} have also used a private dataset containing 660 hours of ECG data collected from 43 participants. Lin et al.  \cite{dp015} used a private dataset from wearable sensors where they have collected ECG data of 2.2 millions data points. Heart disease dataset is also used in the research \cite{dp018}. Mohammad et al.  \cite{dp017} have generated a private dataset of heart rates from wearable devices which were attached to a user for two weeks. On the other hand  Guan et al.  \cite{dp033} have utilised blood record dataset in their research.The blood dataset contains individual information of blood donation and the other dataset contains the identity of the individuals and other general information.

In addition, many researchers have worked with activity type data such as walking, running, sleeping.  Kim et al.  \cite{dp058} have prepared a dataset for daily step counts collected using Gear S3 smartwatch between a limited time period and then replicated 10, 100 \& 1000 times. In another work, Kim et al.  \cite{dp059} have utilised a private data set consisting of daily cumulative step-count data of 247 days where each cumulative step-count data corresponds a stream of length 600. Kim et al.  \cite{dp419} have also used a public PAMAP2 physical activity monitoring data collected from  \cite{dp725} which contains a heart rate monitoring dataset that is collected using sensors. To track physiological activity, smart-home data is also used \cite{dp711, dp712}. Data are collected under 7 scenarios, including sleeping, resting, dressing, eating, toilet use,hygiene and communication.  

Bozkir et al.  \cite{dp521}  also used two public datasets: i) public eye-tracking dataset collected with an Oculus VR device  and ii) pupil eye-tracking dataset from  \cite{dp732} where 20 participants were tasked with reading three different document types (a comic, newspaper, and textbook) in a VR environment. The utilisation of different datasets is summarised in Table \ref{tab:rq3Physio}.

\begin{table*}[t]
\centering
\caption{Different types of datasets}
\label{tab:rq3Physio}
\begin{tabular}{l|l|cccc}
\hline
\rowcolor[gray]{0.9} 
\multicolumn{2}{l|}{\cellcolor[gray]{0.96}\textbf{Data Type}} & \textbf{Availability} & \textbf{Physiological} & \multicolumn{1}{l}{\cellcolor[gray]{0.96}\textbf{Real-time}} & \multicolumn{1}{l}{\cellcolor[gray]{0.96}\textbf{Others}} \\ \hline \hline
\multicolumn{2}{l|}{} & Private & \cite{dp015}, \cite{dp017}, \cite{dp019}, \cite{dp482} & - & - \\ 
\multicolumn{2}{l|}{\multirow{-2}{*}{Heart-related}} & \cellcolor[gray]{0.96}Public & \cellcolor[gray]{0.96}\cite{dp018}, \cite{dp139} & \cellcolor[gray]{0.96}- & \cellcolor[gray]{0.96}- \\ \hline
\multicolumn{2}{l|}{Blood dataset} & Public & \cite{dp717} & - & - \\ \hline
\multicolumn{2}{l|}{} & \cellcolor[gray]{0.96}Private & \cellcolor[gray]{0.96}\cite{dp058}, \cite{dp059} & \cellcolor[gray]{0.96}- & \cellcolor[gray]{0.96}- \\ 
\multicolumn{2}{l|}{\multirow{-2}{*}{\begin{tabular}[c]{@{}l@{}}Activity (Step Count,\\  Running)\end{tabular}}} & Public & \cite{dp419} & - & - \\ \hline
\multicolumn{2}{l|}{Eye tracking} & \cellcolor[gray]{0.96}Public & \cellcolor[gray]{0.96}\cite{dp521} & \cellcolor[gray]{0.96}- & \cellcolor[gray]{0.96}- \\ \hline
 & \begin{tabular}[c]{@{}l@{}}Wearable  sensors\end{tabular} & Private & \cite{dp016}, \cite{dp015}, \cite{dp060} & - & - \\ 
 & \cellcolor[gray]{0.96}Smart home & \cellcolor[gray]{0.96}Public & \cellcolor[gray]{0.96}\cite{dp444} & \cellcolor[gray]{0.96}\cite{dp040} & \cellcolor[gray]{0.96}- \\ 
 & \begin{tabular}[c]{@{}l@{}}GPS, Traffic,  Weight\end{tabular} & Public & \multicolumn{1}{c}{-} & \multicolumn{1}{c}{\cite{dp160}, \cite{dp666}} & \multicolumn{1}{c}{-} \\ 
 & \cellcolor[gray]{0.96}\begin{tabular}[c]{@{}l@{}}Flu, Traffic and\\ Unemployment\end{tabular} & \cellcolor[gray]{0.96}Public & \multicolumn{1}{c}{\cellcolor[gray]{0.96}-} & \multicolumn{1}{c}{\cellcolor[gray]{0.96}\cite{dp707}, \cite{dp020}, \cite{dp688}} & \multicolumn{1}{c}{\cellcolor[gray]{0.96}-} \\ 
& \begin{tabular}[c]{@{}l@{}}Microsoft Band and \\Community Health\end{tabular} & Private & \multicolumn{1}{c}{-} & \multicolumn{1}{c}{-} & \multicolumn{1}{c}{\cite{dp469}} \\ 
\multirow{-6}{*}{Mix} & \cellcolor[gray]{0.96}  Mobile health & \cellcolor[gray]{0.96} Public & \multicolumn{1}{c}{\cellcolor[gray]{0.96}-} & \multicolumn{1}{c}{\cellcolor[gray]{0.96}-} & \multicolumn{1}{c}{\cellcolor[gray]{0.96}\cite{dp442}\cellcolor[gray]{0.96}} \\ \hline
\end{tabular}
\end{table*}

\vspace{2mm}
\noindent \textbf{Real-time category:} Real-time data publishing is also validated by using datasets. Liyue et al.  \cite{dp020} have utilised three different sets of public datasets, namely Flu dataset  \cite{dp713}, Traffic dataset  \cite{dp715} and  Unemployment dataset  \cite{dp716} and tried to correlate them. The same Flu dataset has been harnessed by Gao et al.  \cite{dp688} as well. Rastogi et  al. \cite{dp160} have utilised different public datasets such as GPS, Traffic  and Weight where the last dataset contained daily weight data of about 300 users. Fan et al.  \cite{dp707} have utilised three real-world public datasets which are Flu  \cite{dp713}, Unemployment  \cite{dp716} and Traffic  \cite{dp724}.  Wang et al.  \cite{dp666} experimented with two real-world public datasets, Taxi Trajectory Prediction  \cite{dp733} and World Cup  \cite{dp734} and one synthetic Spatio-temporal dataset called Brinkhoff  \cite{dp735}. Utilised datasets for the real-time category are summarized in Table \ref{tab:rq3Physio}.

\vspace{2mm}
\noindent \textbf{Others category:} Yang et al.  \cite{dp442} have used two real-world public datasets, namely Reference Energy Disaggregation Dataset and Mobile Health Dataset  which consists of 1 million records from 24 different sensor signals. Luo et al.  \cite{dp469} also have used two real-world private case studies. The first one was a community health survey consisting of the heart rate of 20 students in order to find the average heart rate and heart rate distribution. The second dataset was a collaborative emotion classification dataset in which Microsoft Band was used to collect the heart rate, GSR, and skin temperature from users to build collaborative classification models. We have summarised the used datasets in the others category in Table \ref{tab:rq3Physio}.

\begin{figure}[ht]
    \includegraphics[width=0.45\textwidth, center]{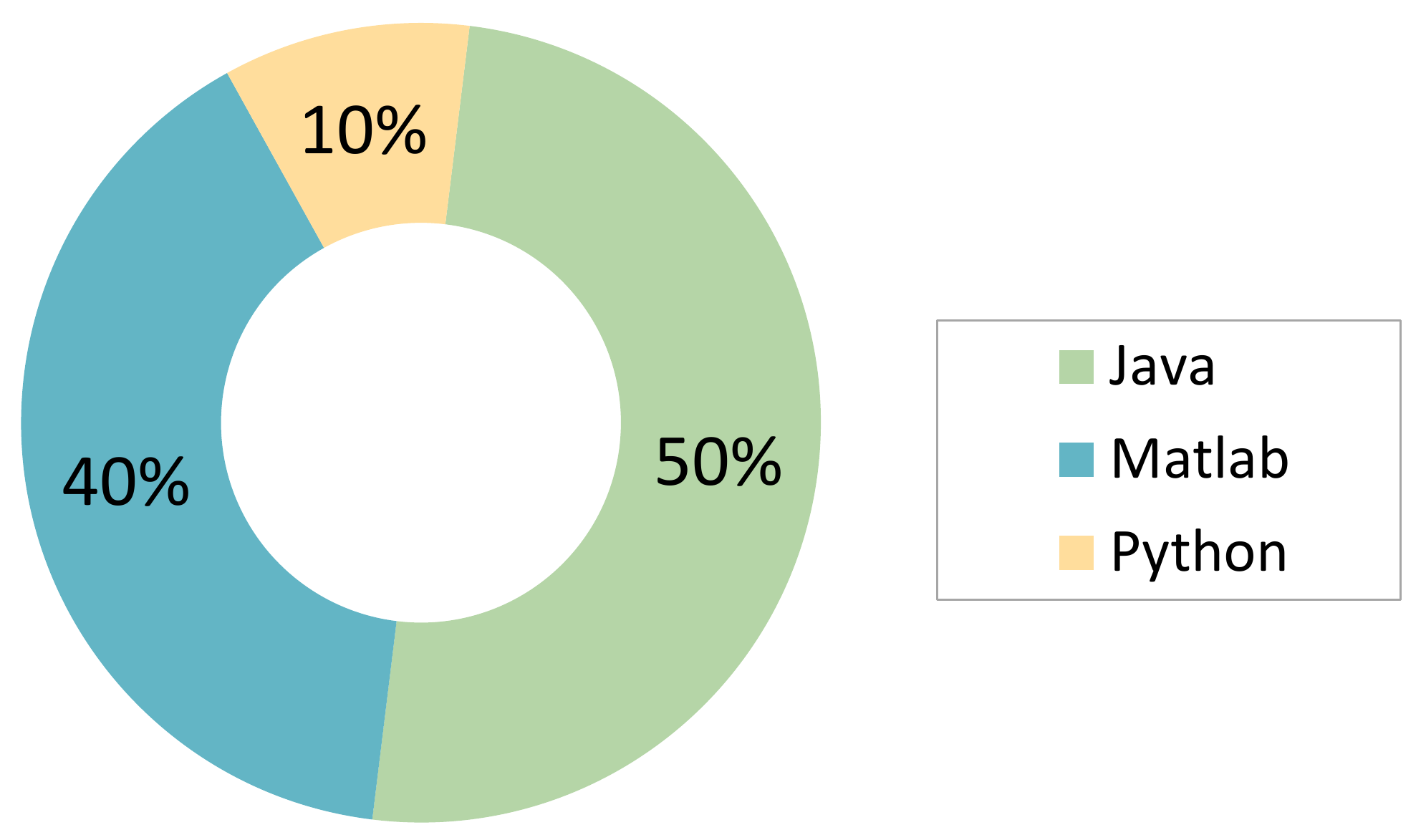}
    \caption{Languages used for perturbation schemes}
    \label{fig_language}
\end{figure}

\vspace{2mm}
\noindent \textbf{Programming languages:}
In addition to dataset, we have also reviewed what programming languages have been used to implement differential privacy. In our survey, we have found that \textit{Java}. \textit{MATLAB},  and \textit{Python} are frequently used by researchers for implementing their algorithms. Fig. \ref{fig_language} shows how different programming languages are used for implementation. However, recently python has emerged as the most used programming language  in terms of open source differential privacy implementation \cite{github2020}. Google has also open sourced their differential privacy library \cite{Vincy2019}, which is implemented using C++ language.

\subsection{RQ4: What are the privacy criteria used in data publishing?}
Different researchers have considered different privacy criteria for preserving privacy of data. Most common criteria is $\epsilon$-differential privacy($\epsilon$-DP) where $\epsilon$ is the privacy budget associated with any data release. Other than this, for real-time data, Shi et al  \cite{dp253} have provided a methodology that depends on ($\epsilon$, $\delta$)-differential privacy(($\epsilon$, $\delta$)-DP). At the same time some have considered new privacy criterion such as w-event and w-day $\epsilon$-differential privacy  \cite{dp040, dp666, dp709}. In these mechanisms, privacy budget is calculated for w consecutive events. Liyue et al.  \cite{dp020} have solved the privacy preserving problem a bit differently. Their proposed solution satisfies $\alpha/T$-differential privacy, where $\alpha$ is the privacy budget and $T$ is the length of the entire series. Wang et al.  \cite{dp040} have collected health data for $w$ different days and preserved privacy for those consecutive days. 
We have presented a summary of different privacy criteria in Table \ref{tab:privacy_criteria}.

\begin{table}[ht]
\caption{Different privacy criterion adopted by researchers}
\centering
\label{tab:privacy_criteria}
\begin{tabular}{c|ll}
\hline
\rowcolor[gray]{.9} 
\multicolumn{1}{l|}{\textbf{Category}} & \textbf{Privacy Criterion} & \textbf{List of Papers}                         \\ \hline \hline
\multirow{2}{*}{Physiological} &
  $\epsilon$-DP &
  \begin{tabular}[c]{@{}l@{}} \cite{dp018}, \cite{dp017}, \cite{dp444}, \cite{dp422}, \\\cite{dp026}, \cite{dp033}, \cite{dp058}, \cite{dp059}, \\\cite{dp419}, \cite{dp060}, \cite{dp140}, \cite{dp482}, \\\cite{dp521}, \cite{dp139}\end{tabular} \\  
                                        & 
                                        \cellcolor[gray]{.96}
                                        $w$-day $\epsilon$-DP           & \cellcolor[gray]{.96} \cite{dp019}                              \\ \hline
\multirow{5}{*}{Real-time} &
  $\epsilon$-DP &
  \begin{tabular}[c]{@{}l@{}} \cite{dp160},  \cite{dp437},  \cite{dp688}\end{tabular}  \\ 
                                        & \cellcolor[gray]{.96} ($\epsilon$, $\delta$)-DP        & \cellcolor[gray]{.96}  \cite{dp253}                                \\  
                                        & 
                                        $w$-event $\epsilon$-DP         & \cite{dp040}, \cite{dp666},  \cite{dp709}  \\  
                                        & \cellcolor[gray]{.96} $(\alpha/T)$-DP               & \cellcolor[gray]{.96} \cite{dp020}                              \\  
                                        &
                                        $\alpha$-DP                   & \cite{dp707}                                \\ \hline
                                        
\multicolumn{1}{l|}{Others}           & \cellcolor[gray]{.96} $\epsilon$-DP                 & \cellcolor[gray]{.96} \cite{dp427},  \cite{dp442}            \\ \hline
\end{tabular}
\end{table}

\subsection{RQ5: What are the limitations of the proposed solutions?}

Researchers have conducted several experiments for evaluating their proposed methodology. However, these experiments have several limitations. The limitations of existing methodologies need to be addressed for a suitable, and more practical privacy preserved framework.

\vspace{2mm}
\noindent \textbf{Physiological category:} For physiological data, the proposed DP solutions mainly suffer from  scalability issues \cite{dp482}.  Many of the proposed differential privacy models are strict to static data publishing and confined to single dimension \cite{dp422}. Moreover, many of the privacy protection schemes are just theoretical in nature \cite{dp439, dp016}. Some models \cite{dp444} suffer from performance degradation with an increasing number of cloud resources. The method in  \cite{dp026} introduced relative errors with a heavy computational burden on cloud servers. Finally, the algorithm proposed by Mohammad et al.  \cite{dp017} has a higher complexity than existing works (e.g. Li's  \cite{ref_mohammad_li}).
In addition, the proposed methods also vulnerable to information leakage in the presence of a strong adversarial model. It can cause the adversary gain more knowledge and result in privacy leakage.

\vspace{2mm}
\noindent \textbf{Real-time category:} For real-time data, the proposed DP solutions mainly suffer from different types of errors such as reconstruction \& perturbation errors  \cite{dp160}, relative errors  \cite{dp707, dp040} and absolute error  \cite{dp688} have been observed. To detail, an algorithm's accuracy is found to be affected by failures occurred when the algorithm is executing to answer a query  \cite{dp160}. For non-linear synthetic datasets, the proposed method has a higher relative error due to a model misfit compared to the existing methods  \cite{dp707}. One of the most common problem is the difficulty choosing an optimal value for epsilon ($\epsilon$) to gain any advantage. For example, for a large budget ($\epsilon > 1$), there is no substantial advantages \cite{dp020} and even some proposed method under perform with a higher epsilon value  \cite{dp040}. Similarly, in  \cite{dp688}, it has been found that increasing in epsilon value has weakened the algorithm. Thus, choosing an appropriate value for threshold is a challenging task. Besides,the absolute error in the GGA algorithms is smaller than other algorithms only when the query range is greater than 40. Authors in  \cite{dp253}, have noted a number of issues such as dynamic join and aggregation problem. Also, when a node failure occurs, some of the participants become unable to provide their encrypted data. Another limitation is lack of support for graceful degradation during node failures. 

\vspace{2mm}
\noindent \textbf{Others category:} As for the others category, communication overheads with estimation errors have been reported as the major shortcoming  \cite{dp427}. In  \cite{dp459}, local and global errors have been highlighted, compared to the Laplace mechanism, because of the usage of sine polyonym.  System and computational overheads have incurred due to usage of Salus in  \cite{dp469}. In addition, computational overhead and data reconstruction error have been experienced. Maintaining a balanced noise and sensitivity have also been an obstacle in the way of preserving privacy as large training sets contain too much noise which results in the loss of utility for the proposed model  \cite{dp442}.

\vspace{2mm}
\noindent \textbf{Summary:} Different types of limitations found in different approaches are summarized in Table \ref{tab:categorized_limitations}.

\begin{table*}[t]
\centering
\caption{Limitations in the current research works}
\label{tab:categorized_limitations}
\begin{tabular}{l|l|lll}
\hline
\rowcolor[gray]{.9}
\multicolumn{1}{l|}{\textbf{Limitations}} &
  \textbf{Type} &
  \multicolumn{1}{l}{\textbf{Physiological}} &
  \multicolumn{1}{l}{\textbf{Real-time}} &
  \multicolumn{1}{l}{\textbf{Others}} \\ \hline \hline
\multirow{7}{*}{Error} &
  
  Relative error &
  - &
  \multicolumn{1}{l}{\cite{dp707}, \cite{dp026}} &
  - \\ 
 &
 \cellcolor[gray]{.96}
  Reconstruction error & \cellcolor[gray]{.96}
  - & 
  \multicolumn{1}{l}{\cellcolor[gray]{.96} \cite{dp160}} & \cellcolor[gray]{.96}
  - \\ 
 &
 
  Perturbation error &
  - &
  \multicolumn{1}{l}{\cite{dp160}} &
  - \\ 
 &
 \cellcolor[gray]{.96}
  Absolute error & \cellcolor[gray]{.96}
  - & 
  \multicolumn{1}{l}{\cellcolor[gray]{.96}\cite{dp688}} & \cellcolor[gray]{.96}
  - \\ 
 &
  \begin{tabular}[c]{@{}l@{}}Error increased due to \\ larger group\end{tabular} &
  - &
  \multicolumn{1}{l}{\cite{dp709}} &
  - \\ 
 &
 \cellcolor[gray]{.96}
  Estimation error & \cellcolor[gray]{.96}
  - & \cellcolor[gray]{.96}
  - &
  \multicolumn{1}{l}{\cellcolor[gray]{.96}\cite{dp427}} \\ 
 &
  Data reconstruction error &
  - &
  - &
  \multicolumn{1}{l}{\cite{dp469}} \\ \hline
\multirow{3}{*}{Overhead} &
    \cellcolor[gray]{.96}
  Communication Overhead & \cellcolor[gray]{.96}
  - & \cellcolor[gray]{.96}
  - &
  \multicolumn{1}{l}{\cellcolor[gray]{.96}\cite{dp427}} \\ 
 &
  Computational overhead &
  \multicolumn{1}{l}{\cite{dp026}} &
  - &
  \multicolumn{1}{l}{\cite{dp469}} \\ 
 &
 \cellcolor[gray]{.96}
  System Overhead & \cellcolor[gray]{.96}
  - & \cellcolor[gray]{.96}
  - & 
  \multicolumn{1}{l}{\cellcolor[gray]{.96}\cite{dp469}} \\ \hline
\begin{tabular}[c]{@{}l@{}}Appropriate value \\ for epsilon \end{tabular}&
  \multicolumn{1}{c|}{-} &
  \multicolumn{1}{l}{\cite{dp016}} &
  \multicolumn{1}{l}{\begin{tabular}[c]{@{}l@{}}\cite{dp020}, \cite{dp040}, \\ \cite{dp688}, \cite{dp707}, \cite{dp709}\\  \end{tabular}} &
  \multicolumn{1}{l}{\cite{dp442}} \\ \hline
\multirow{3}{*}{Algorithm} &
  \cellcolor[gray]{.96}
  Complexity &
  \multicolumn{1}{l}{\cellcolor[gray]{.96}\cite{dp017}} & \cellcolor[gray]{.96}
  - & \cellcolor[gray]{.96}
  - \\ 
 &
  \begin{tabular}[c]{@{}l@{}}Unable to detect \\ tempered data\end{tabular} &
  \multicolumn{1}{l}{\cite{dp140}} &
  - &
  - \\ 
 &
 \cellcolor[gray]{.96}
  Scalability & 
  \multicolumn{1}{l}{\cellcolor[gray]{.96}\cite{dp482}} & \cellcolor[gray]{.96}
  - & \cellcolor[gray]{.96}
  - \\ \hline
\end{tabular}
\end{table*}

\section{Discussion}
\label{sec:discussion}

Differential privacy has paved the way for a more flexible solution in privacy preservation. It has overcome the limitations of existing methodologies to some extent. However, the basic differentially private perturbation method alone cannot protect data from getting exposed. Several researchers have identified the major concerns regarding challenges of publishing such data by using basic mechanisms of differential privacy \cite{challenges1, challenges2}. Therefore, researchers have proposed to combine different methods with differential privacy in order to provide an effective privacy protection mechanism. They have tried to propose their mechanisms in such a way that both the privacy and data utilization are well balanced.

Throughout the paper we have explored, through our research questions, the advantages of applying differential privacy over other techniques, challenges faced by basic mechanisms and how different researchers have extended the basic differential privacy to meet the requirements of wearable data publishing. In this section, we have summarised our findings from different perspectives. 

\begin{figure*}[ht]
  \begin{subfigure}{0.49\textwidth}
    \includegraphics[width=\linewidth]{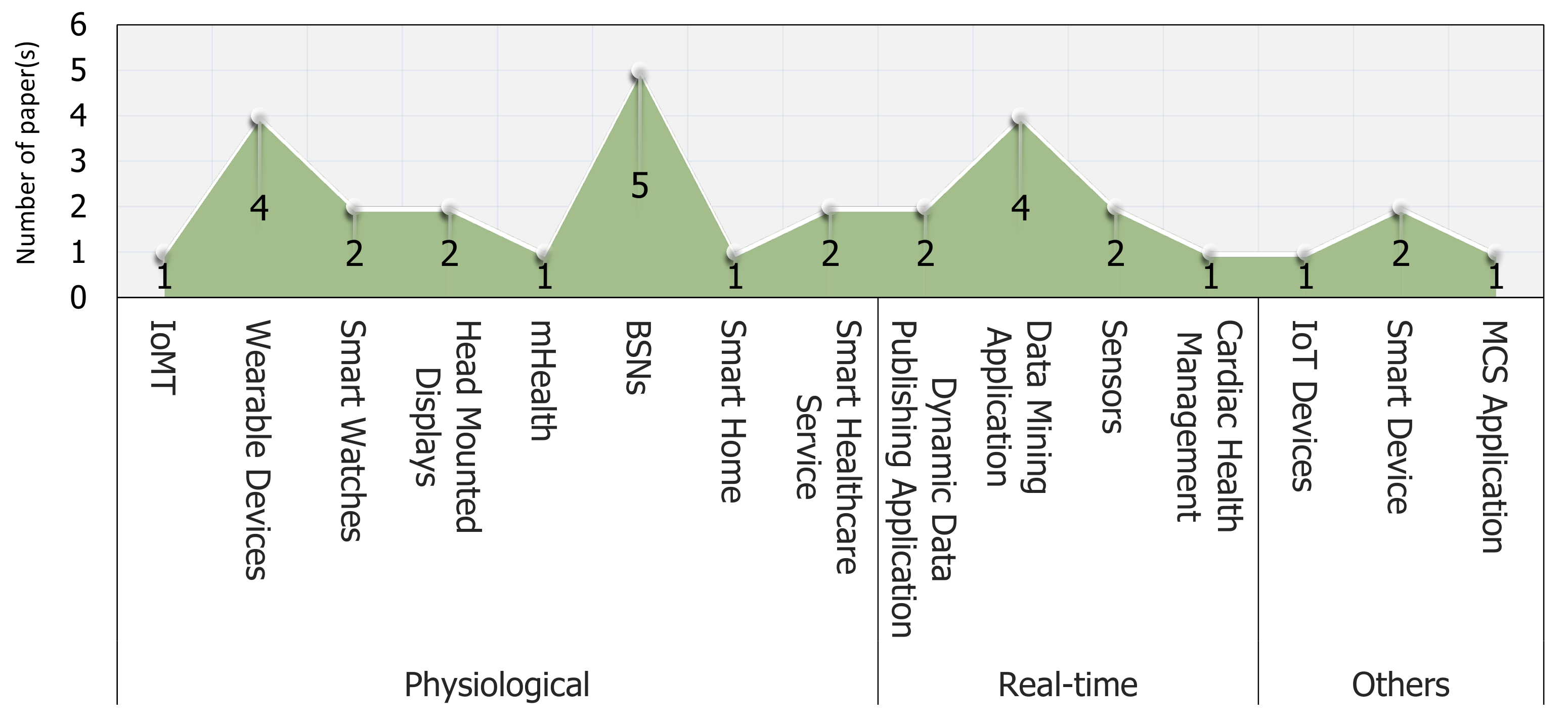}
    \caption{Different application domains} \label{fig_applied_area}
  \end{subfigure}%
  \hspace*{\fill}  
  \begin{subfigure}{0.49\textwidth}
    \includegraphics[width=\linewidth]{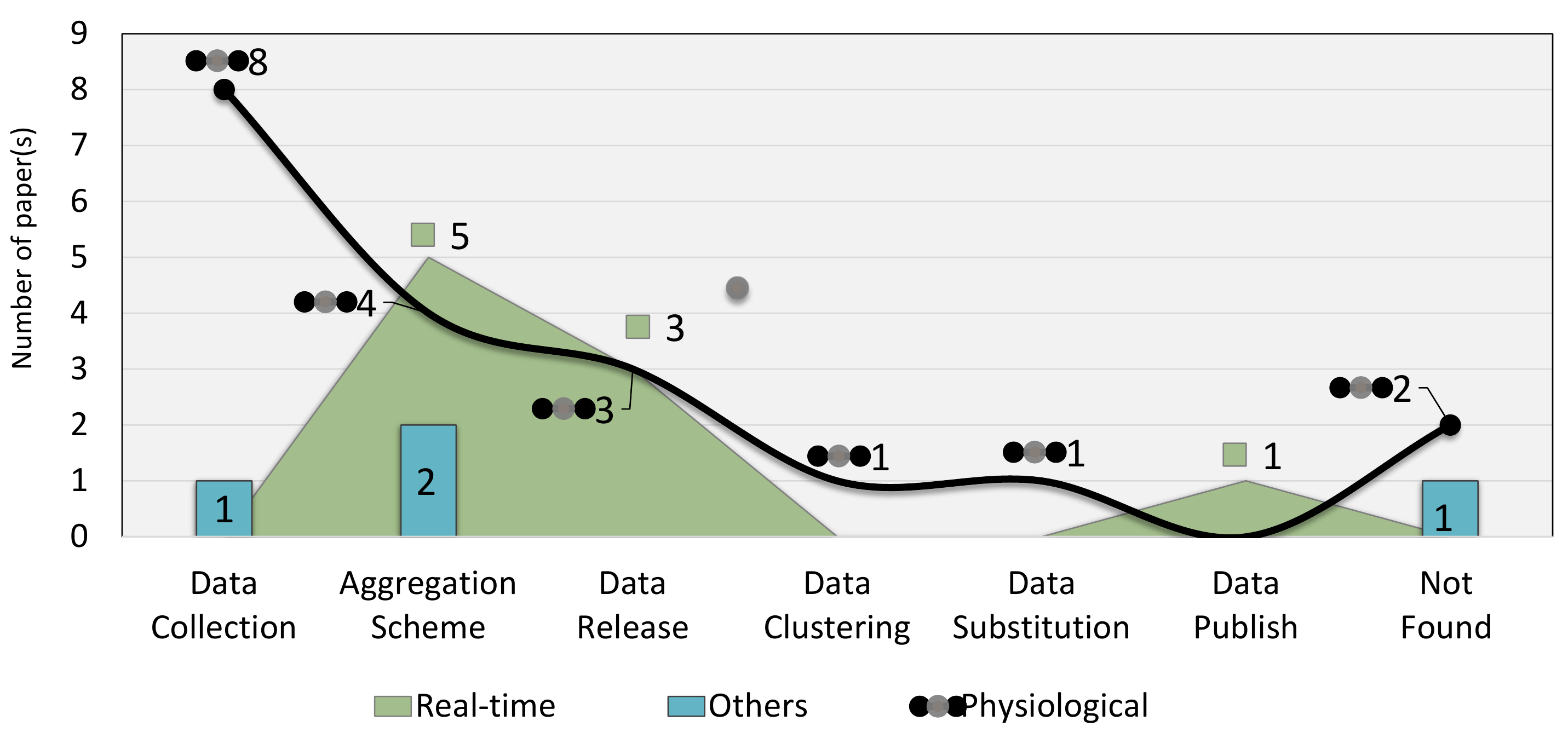}
    \caption{Different data management techniques}  \label{fig_application_criteria}
  \end{subfigure}%
    \caption{Number of research papers in different application domains and data management}
\end{figure*}

\subsection{Research Question Perspective}
At first, we present the following summary from the perspective of our research questions.
    \begin{itemize}
        \item \textbf{Application domains:}
        Researchers have considered different application domains for their proposed mechanisms. We have illustrated the number of published papers for different domains in Fig. \ref{fig_applied_area}. As evident from the figure, most of the researches have explored areas such as BSNs, wearable devices, data mining applications and so on. 
        
        \item \textbf{Application scheme:}
        Versatile application schemes have been considered by researchers while preserving data privacy. Some researchers have considered data publishing scheme whereas some of them have proposed aggregation or data release schemes. Fig. \ref{fig_application_criteria} visualizes the proposed schemes. From the figure, for the physiological category, the highest number of papers (8) are for the data collection scheme. On the other hand, the aggregation scheme has the highest number of papers, with 5 and 2, for the real-time and others category respectively.
        \begin{figure}[!b]
                \includegraphics[width=0.5\textwidth, center]{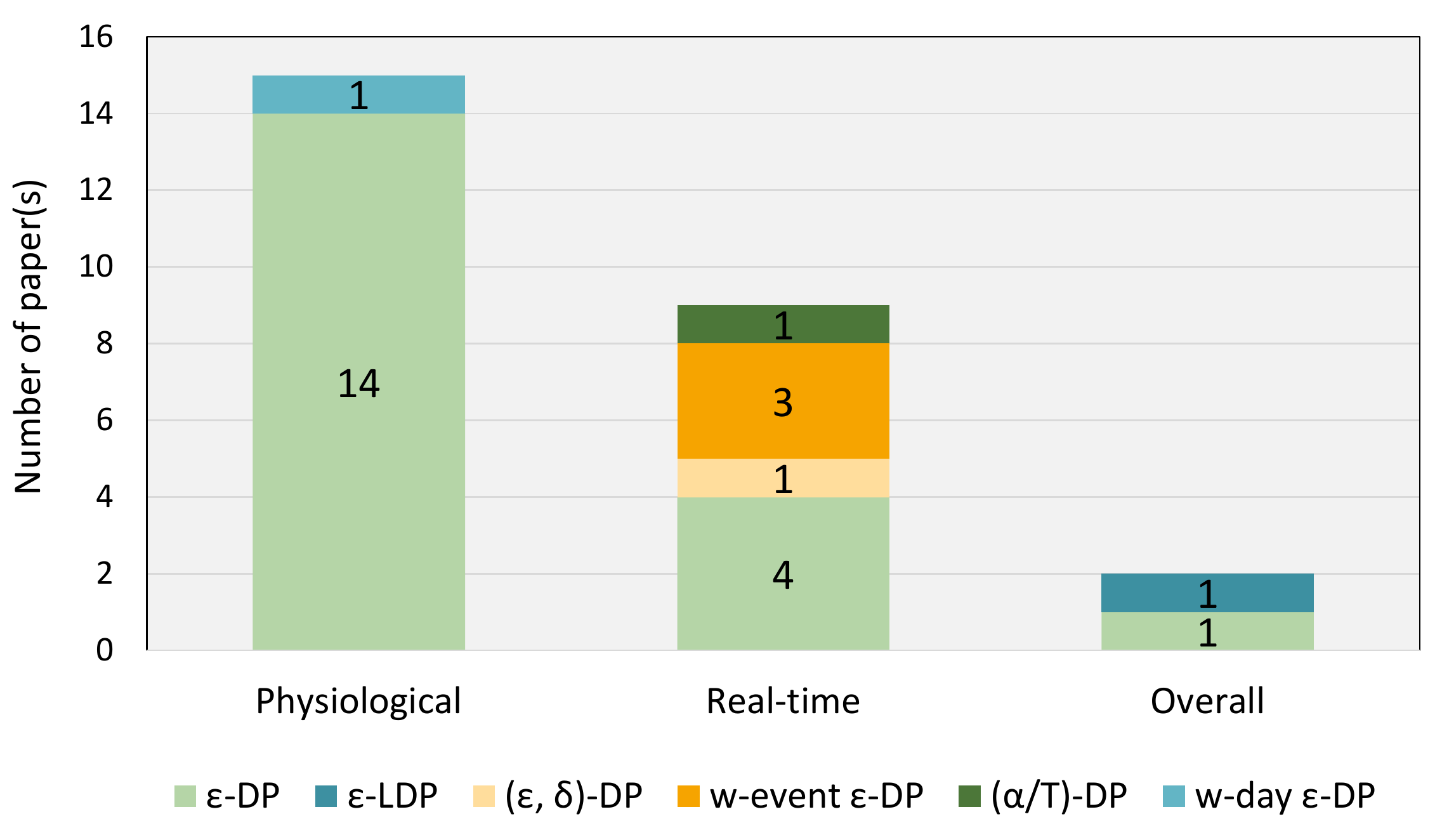}
                \caption{Privacy criteria of proposed schemes}
                \label{fig_privacy_criterion}
            \end{figure}
            \begin{figure}[!b]
                \includegraphics[width=0.5\textwidth, center]{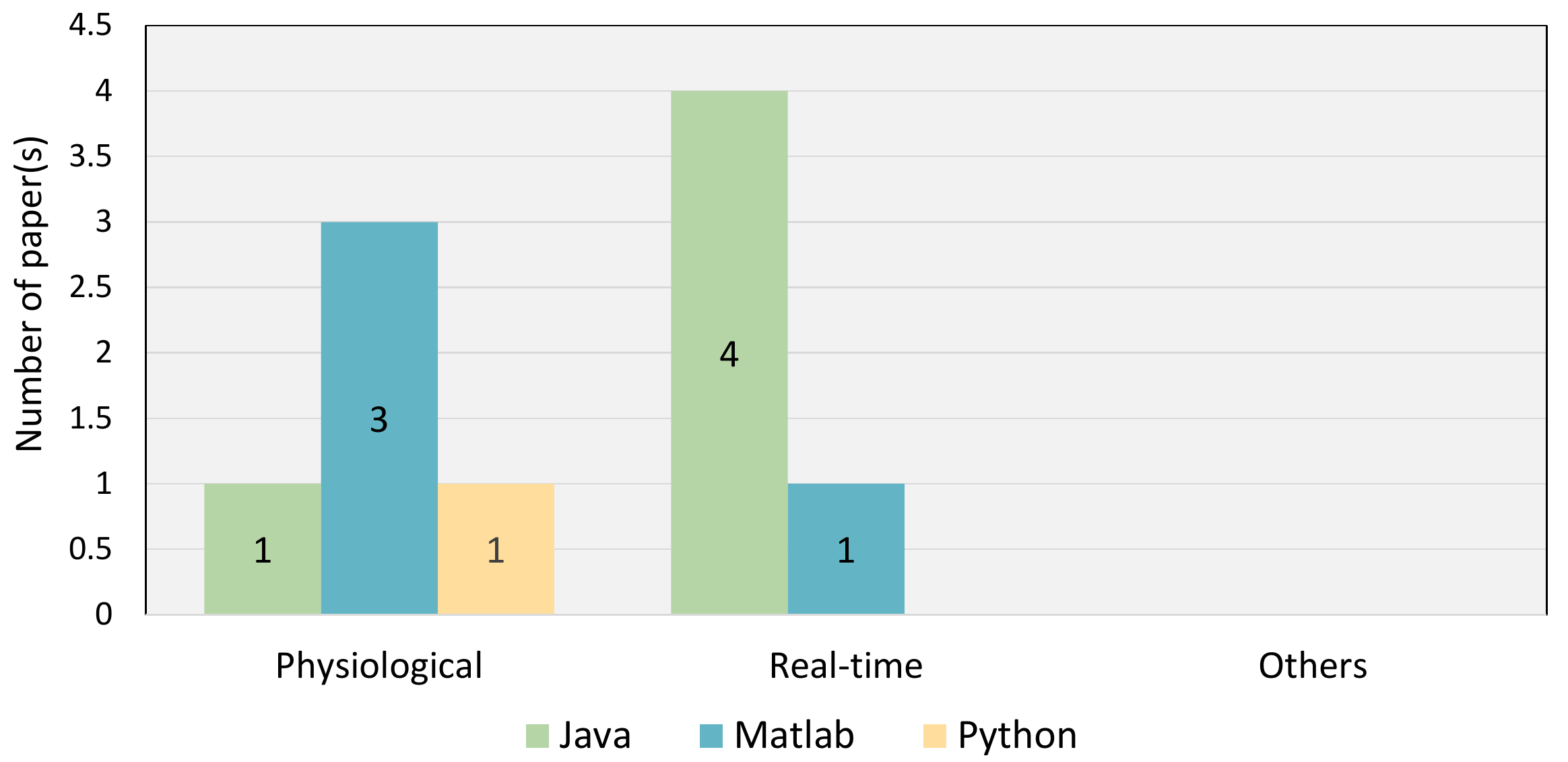}
                \caption{Languages used for developing mechanisms}
                \label{fig_mechanism_developed_language}
            \end{figure}

        \item \textbf{Privacy criteria:}
        In case of privacy criteria, researchers have mostly considered $\epsilon$-DP. Other than this, they have also considered ($\epsilon$, $\delta$)-DP, $\alpha/T$-DP. Some of them even considered $w$-event DP where data is collected for consecutive w-events. Fig. \ref{fig_privacy_criterion} illustrates privacy criteria of proposed schemes in each category: physiological, real-time and others.

        \item \textbf{Programming language used:}
        For implementing differential privacy algorithms, different programming languages such as \textit{Java}, \textit{Python}, \textit{Matlab} have been considered. We have presented a detailed summary of programming languages used in developing the mechanisms in Fig. \ref{fig_mechanism_developed_language}.  

    \begin{figure*}[ht]
  \begin{subfigure}{0.33\textwidth}
    \includegraphics[width=\linewidth]{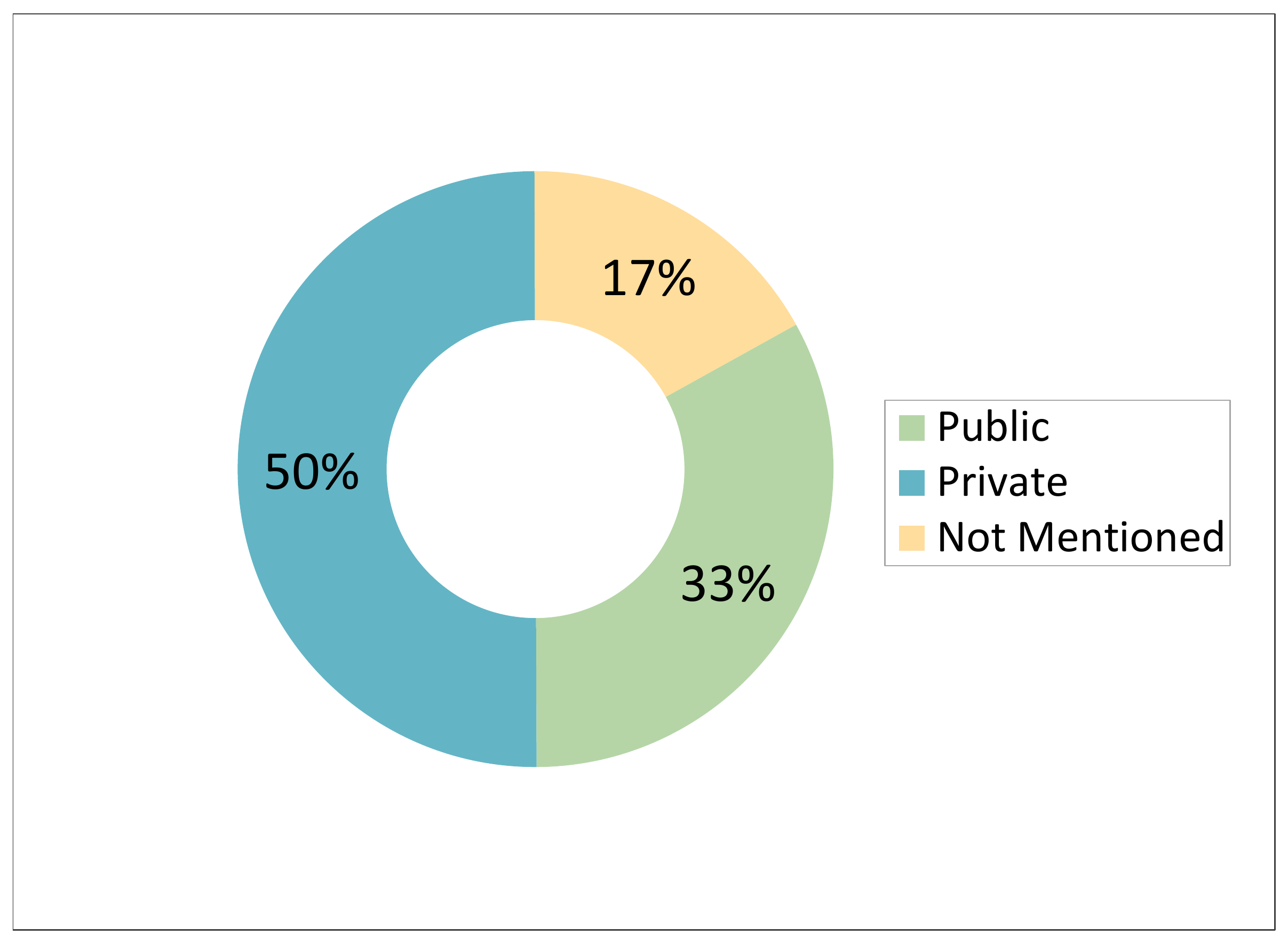}
    \caption{Physiological} \label{Fig:Physio_dataset}
  \end{subfigure}%
  \hspace*{\fill}  
  \begin{subfigure}{0.33\textwidth}
    \includegraphics[width=\linewidth]{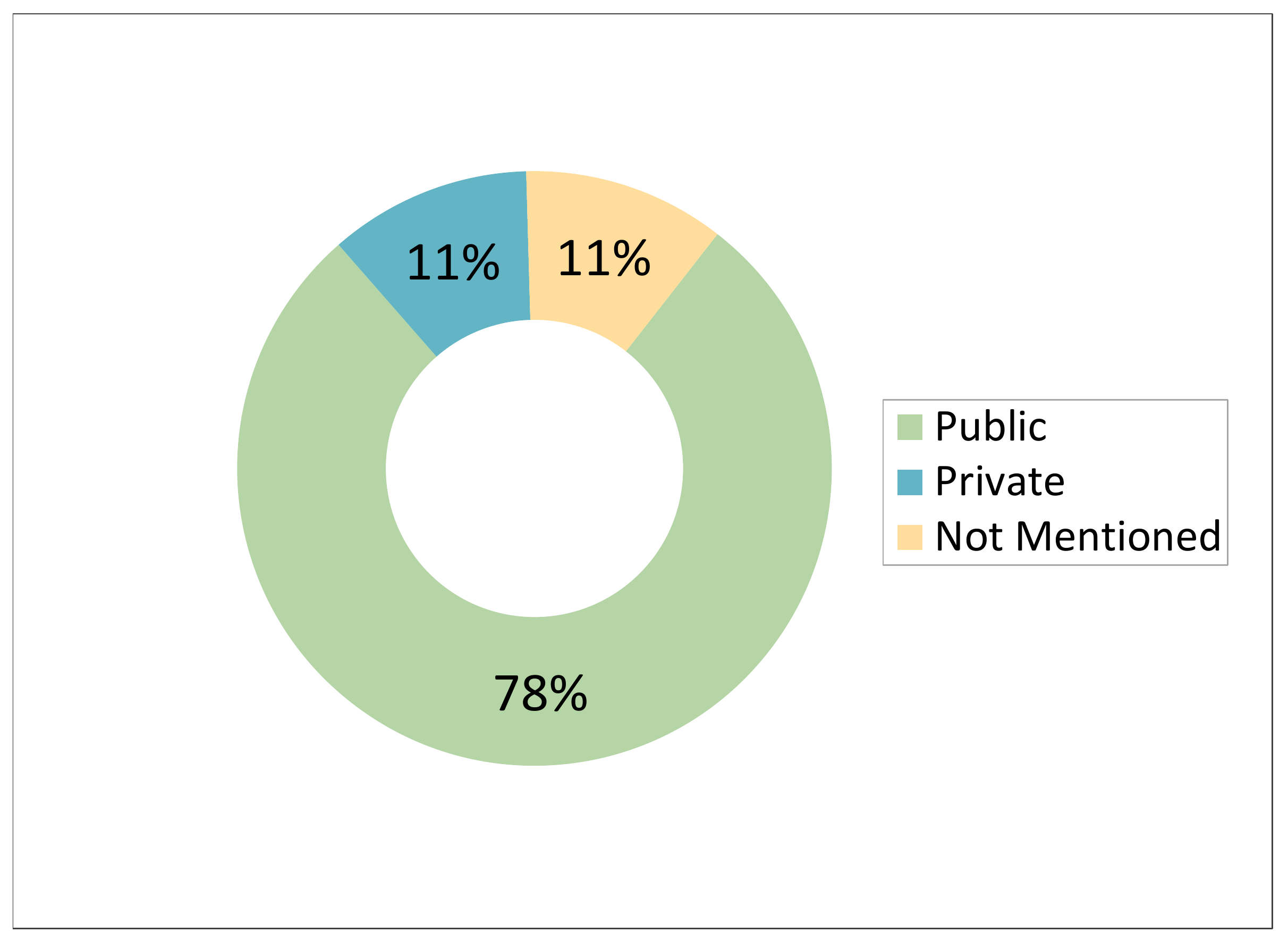}
    \caption{Real-time}  \label{Fig:Realtime_dataset}
  \end{subfigure}%
  \hspace*{\fill}  
  \begin{subfigure}{0.33\textwidth}
    \includegraphics[width=\linewidth]{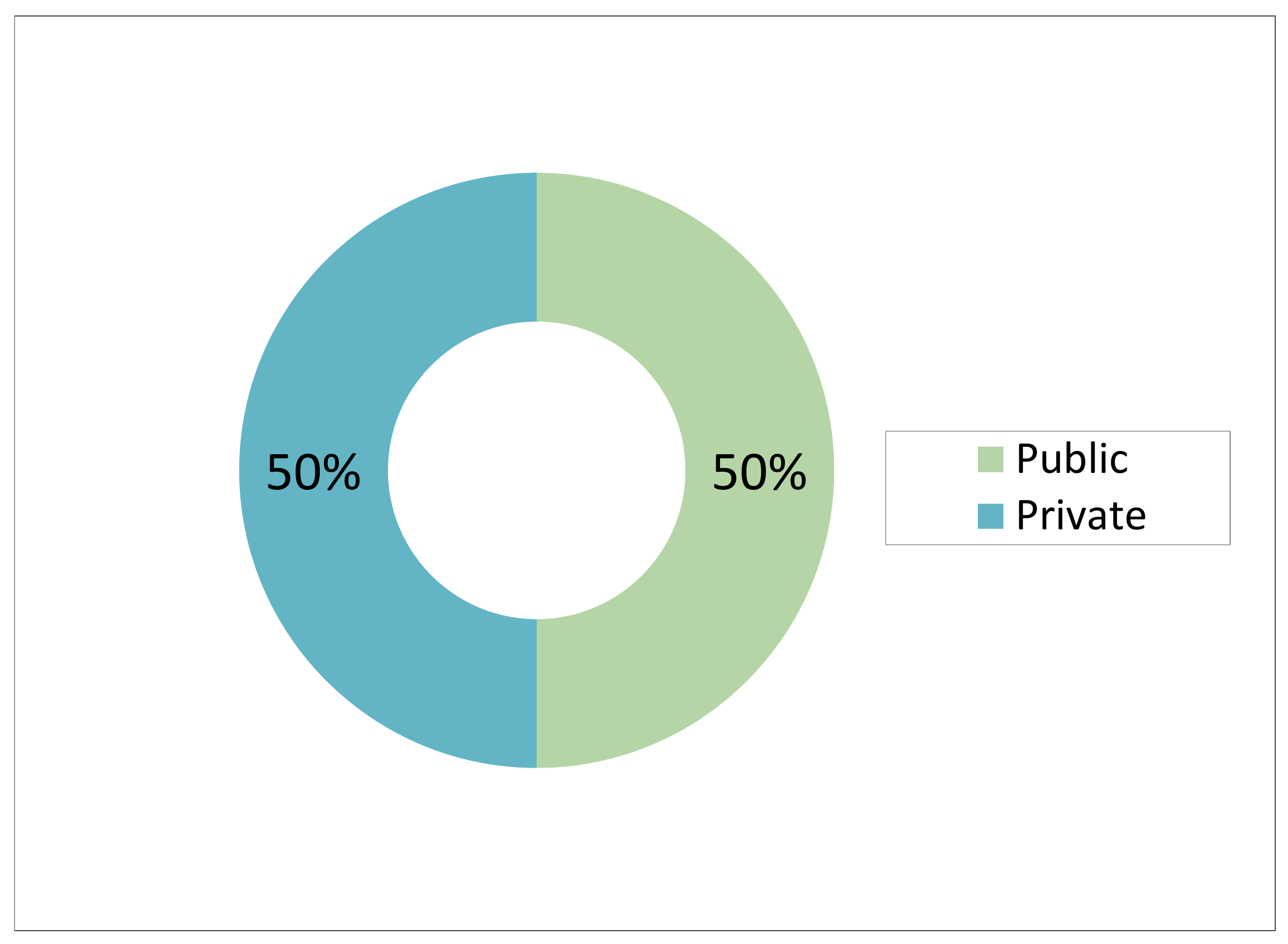}
    \caption{Others}  \label{Fig:Overall_dataset}
  \end{subfigure}
  \caption{Type of datasets used for evaluation} \label{fig:tps}
\end{figure*}    
            
        \item \textbf{Considered datasets for developing mechanisms:}
        After developing the mechanism the most important thing to do is to evaluate the mechanism. For this purpose researchers have used both the private and public datasets. We have provided a summary in Fig. \ref{fig:tps} which visualizes datasets used by different researchers.
           
    \end{itemize} 
\subsection{Data Perspective}

Next, we have summarised all works from the perspective of data: physiological, real-time and others.

\begin{itemize}
    \item \textbf{Physiological data}:
    With the enormous advancement in sensor technology and the  physiological data generated from them, the privacy issues with physiological data is a concerned topic. These sensors generate data which can be dynamically updated or can be temporally correlated. Although, \textit{encryption} and \textit{k-anonymity} are largely adopted solutions for preserving privacy, however, they can be more computationally complex and less practical. Conversely, differential privacy based solutions are more light-weight, more practical and thus have less communication overhead.
    
    18 papers have been selected which have met our selection criteria for the physiological data category. In these 18 papers, researchers have showed various methods for preserving privacy in a better way. Among them, different application behavior such as data collecting, data releasing, aggregation studies for making decisions are majorly noticed. These frameworks have covered areas such as BSNs, wearable devices, smart watches and even in head mounted displays. We have also noticed researchers used Laplace perturbation method most. Which means, numerical queries have been majorly covered. Throughout the method, mostly they have followed $\epsilon$-DP. Other than this, FPA, Gaussian mechanism and $w$-event privacy have also been applied. For conducting experiments, in most cases, they have developed their algorithm in Java and considered their self generated dataset. Fig. \ref{fig_discussion_physiological} makes a summary what we have found among the selected 18 papers.
    
    Researchers have focused on various aspects of physiological data. They have focused mainly on effectiveness of publishing data, how efficiently data can be collected, reducing calculation and communication overhead, maintaining availability and reliability so that proposed solutions provide better accuracy as well as maintain the balance between utility and privacy. Differential privacy based schemes such as, MHDA\textsuperscript{$\oplus$} \cite{dp026}, Re-DPoctor \cite{dp018}, EDPDCS  \cite{dp033}, WSV-MDAV \cite{dp422}, APDP \cite{dp444} outperforms other existing methods. PMHA-DP  \cite{dp140} has less communication overhead than existing solutions. APDP gives more privacy protection in comparison with UDP and NPDP. It also has best performance in terms on attack resistance. The proposed method from \cite{dp521} is capable of handling correlation in data. 

    \begin{figure}
        \centering
        \includegraphics[width=0.47\textwidth]{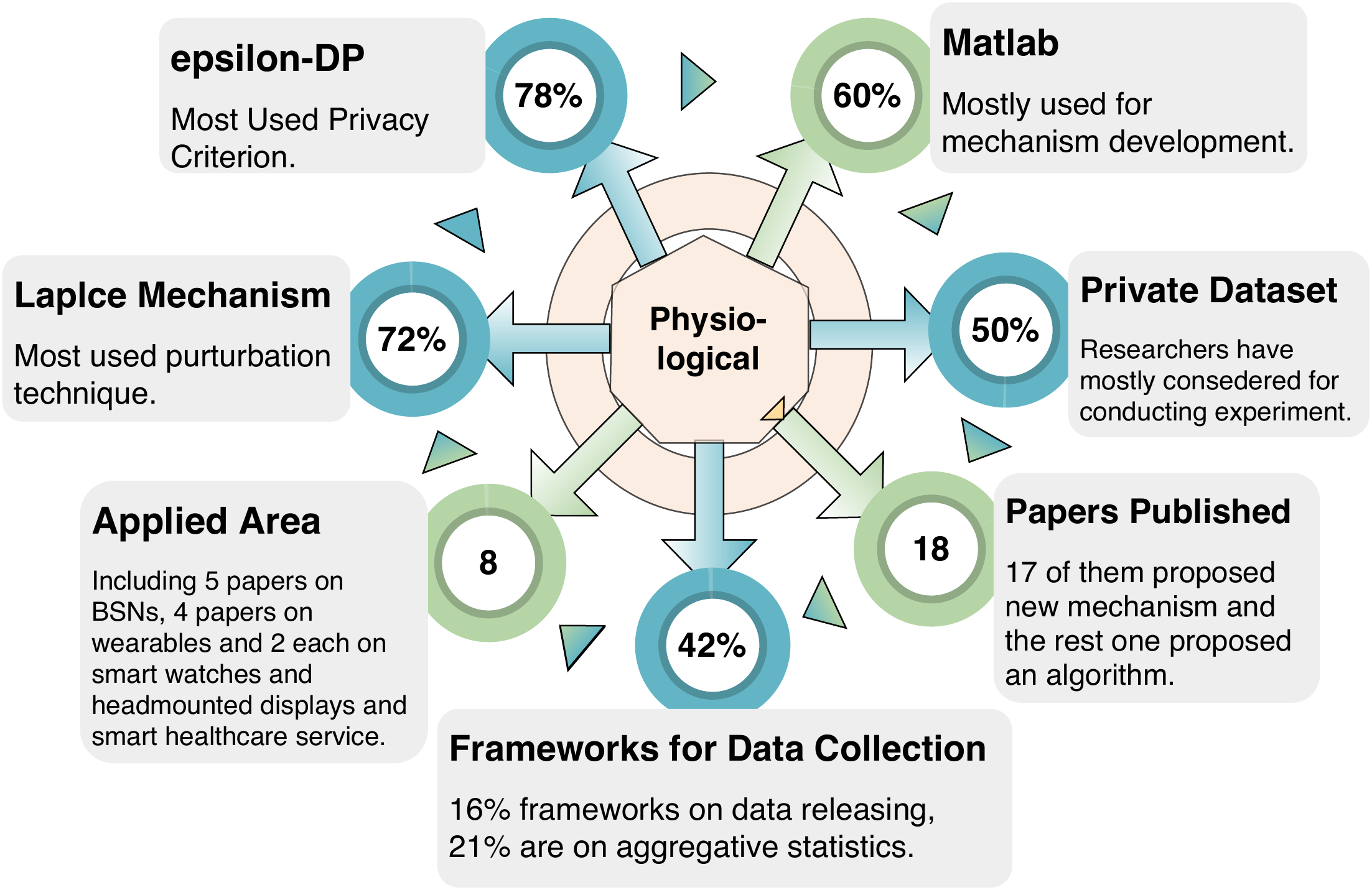}
        \caption{Summary of the reviewed works in the physiological category}
        \label{fig_discussion_physiological}
    \end{figure}
    
    \vspace{0.5cm}
  
   \item \textbf{Real-time data}:
    Differential privacy has also proven itself as a suitable solution in case of preserving privacy of real-time data. We have conducted our review over real-time healthcare sectors as real-time data also generates continuous data. We have found 9 papers satisfying our selection criteria, where majority of researchers have focused in aggregation studies by which data analytics can take important decisions. Other than aggregation, data releasing and publications are barely covered. These frameworks have covered areas such as data mining applications, sensors and dynamic applications as well. 
    
    From our review, we can conclude that the most of the researchers have considered Laplace mechanism as a mean of preserving privacy. Researchers have maintained $\epsilon$-DP for their given mechanisms. Besides, newly adopted privacy criteria such as $w$-event DP, $\alpha/T$-DP has also been observed. In general, developing mechanism in Java and conducting experiment with public dataset is preferred by the researchers. 
    
    For real-time health data, differential privacy based mechanisms outperform existing methods. Researchers have found better results by using differential privacy as a means of privacy protection. Rastogi et al.  \cite{dp160}, Liyue et al.  \cite{dp020}, and Fan et al.  \cite{dp707} have  proposed methods that can achieve better accuracy and utility. The proposal from Rastogi et al. \cite{dp160}, \textit{PASTE}, can also perform excellently under small privacy cost. RescueDP  \cite{dp666}, an aggregate monitoring scheme, outperforms existing solutions as well as improves utility, ensures strong privacy guarantee. Proposed solutions of Wang et al.  \cite{dp040} using Laplace noise with unscented Kalman Filter and Kellaris et al.  \cite{dp709} using Laplace noise with sophisticated sampling and dynamic privacy budget allocation technique are more practical. 
    
    Other mechanisms such as UKFDP  \cite{dp040}, FPA\textsubscript{k}  \cite{dp160}, ConTPL  \cite{dp437} and FAST  \cite{dp707} outperform the existing state-of-the-art methods and provide more practical solutions in terms of accuracy and privacy. Gao et al.  \cite{dp688} have provided a solution that can reduce noise errors and outperform existing solutions. 
    
    In Fig. \ref{fig_discussion_realtime} we have represented a summary for real-time healthcare domain.
    
    \begin{figure}
        \centering
        \includegraphics[width=0.47\textwidth]{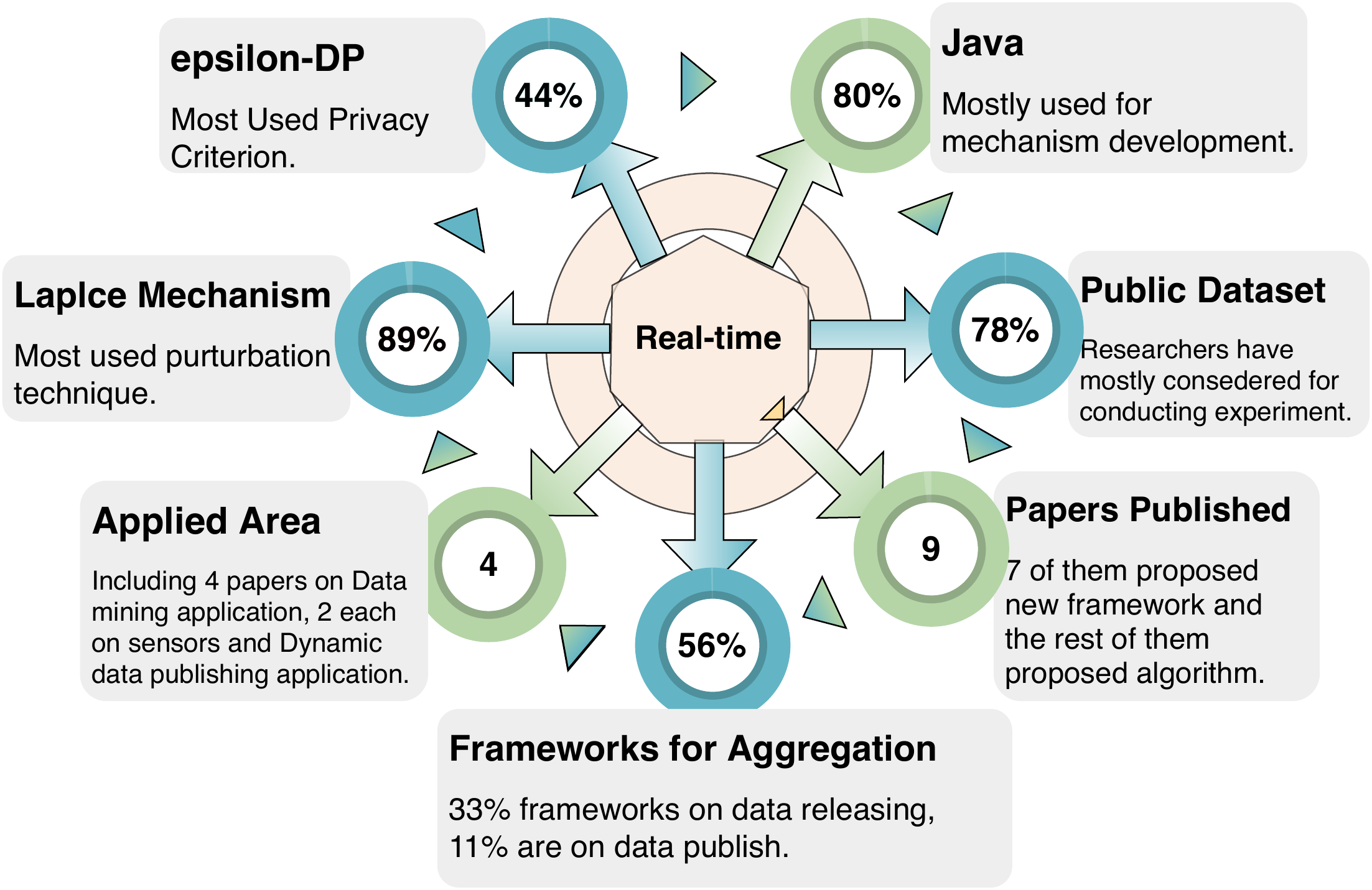}
        \caption{Summary of the reviewed works in real-time category}
        \label{fig_discussion_realtime}
    \end{figure}
    
    \vspace{0.5cm}
    
    \item \textbf{Others}:
     The papers in the others category generate similar patterns like wearable or real-time health data. Differential privacy has also been found beneficial for applying privacy-preserving mechanisms over these data. 
    
    We have reviewed 4 papers which have satisfied our selection criteria. 
    
    The frameworks proposed in these papers cover a wide range of application areas such as smart devices, IoT devices and MCS application. Among all the mechanisms, Laplace mechanism has been mostly adopted by researchers with $\epsilon$-DP being the most widely used privacy criterion. But none of the researchers has mentioned anything about the platform/language they have used for developing either the framework or algorithm. The researchers have conducted various experiments in order to prove the efficiency of their proposed solution and they have used both the public and private datasets equally for their experiments. Among the solutions, Harmony  \cite{dp427} and Salus  \cite{dp469} are more practical, provide accurate and efficient results, reduce errors and maintain a stable balance between privacy and utility by trying to improving accuracy. With the help of fog computing architecture, MLDP \cite{dp442} can aggregate by reducing communication overhead and \cite{dp427} can reduce both the computational and communication overhead. Researchers of \cite{dp469} have claimed that they achieved enhancement in data protection by using differential privacy. In Fig. \ref{fig_discussion_overall}, we have represented a summary for the others category.

    \begin{figure}
        \centering
        \includegraphics[width=0.47\textwidth]{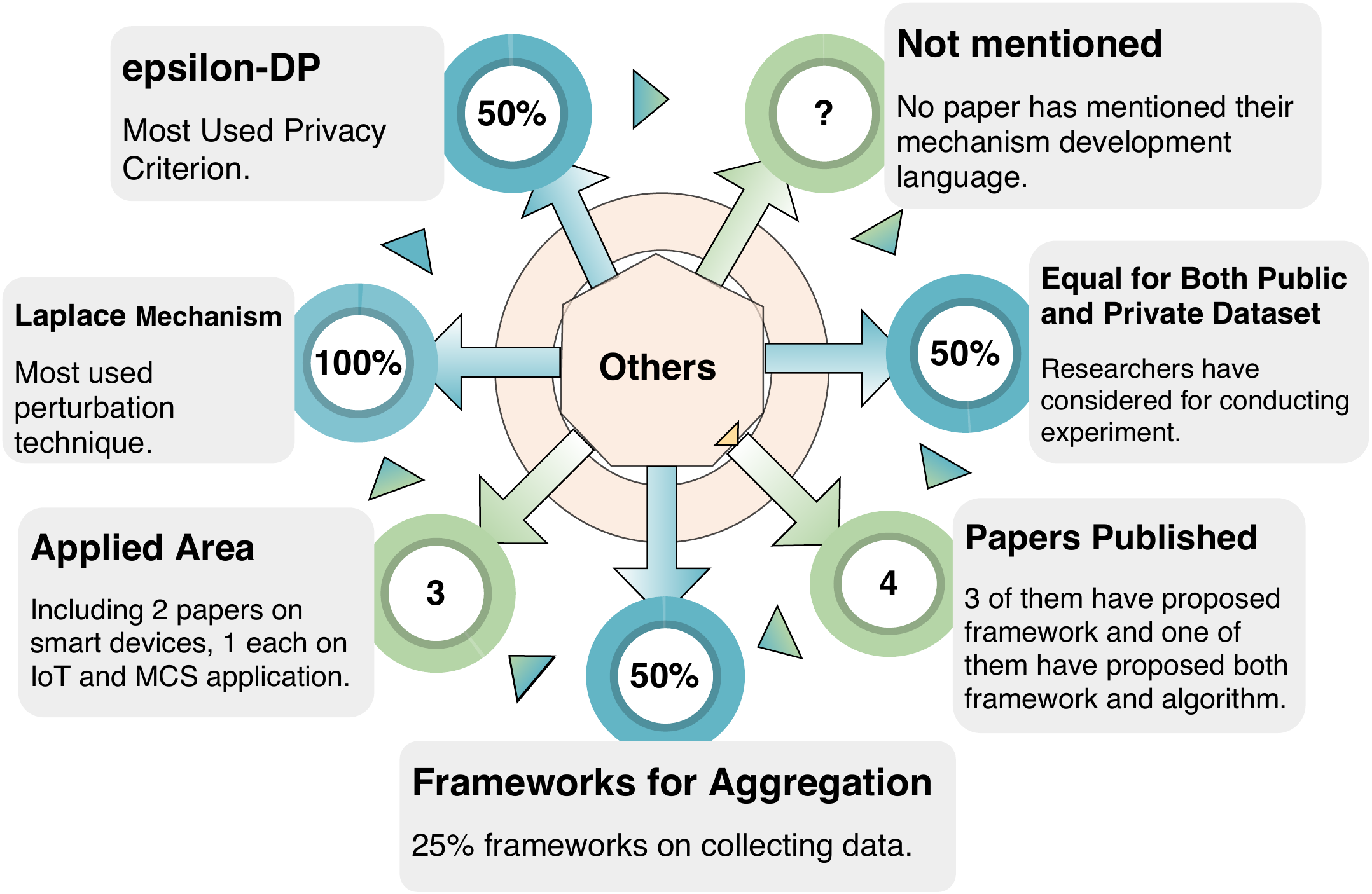}
        \caption{Summary of the reviewed works in the others category}
        \label{fig_discussion_overall}
    \end{figure}

\end{itemize}

\section{Current Challenges and Future Direction}
Table \ref{tab:categorize_major_contribution} provides a summary of the major contribution in terms of wearable data publishing under differential privacy. We have observed that researchers have focused more on utility and performance enhancement compared to security and reliability of the model.

\begin{table}
\centering
\caption{Major contributed area by research communities}
\label{tab:categorize_major_contribution}
\begin{tabular}{l|lll}
\hline
\rowcolor[gray]{0.9}
\textbf{\begin{tabular}[c]{@{}l@{}}Contribution\\ Type\end{tabular}} &
  \textbf{Physiological} &
  \textbf{Real-time} &
  \textbf{Others} \\ \hline \hline
\begin{tabular}[c]{@{}l@{}}Privacy \\ Enhancement\end{tabular} &
  \begin{tabular}[c]{@{}l@{}}\cite{dp018}, \cite{dp015}, \cite{dp422},\\  \cite{dp444}, \cite{dp521}, \cite{dp139}\end{tabular} &
  \cite{dp666} &
  \cite{dp442}, \cite{dp469} \\ 
\rowcolor[gray]{0.96} 
\begin{tabular}[c]{@{}l@{}}Utility \\ Enhancement\end{tabular} &
  \begin{tabular}[c]{@{}l@{}}\cite{dp444}, \cite{dp482}, \cite{dp059},\\  \cite{dp060}
  \end{tabular} &
  \begin{tabular}[c]{@{}l@{}}\cite{dp040}, \cite{dp666}, \\ \cite{dp688}, \cite{dp707}\end{tabular} &
  \begin{tabular}[c]{@{}l@{}}\cite{dp427}, \cite{dp469},\\  \cite{dp459}\end{tabular} \\ 
\begin{tabular}[c]{@{}l@{}}Improved \\ Accuracy\end{tabular} &
  \begin{tabular}[c]{@{}l@{}}\cite{dp016}, \cite{dp017}, \cite{dp033},\\  \cite{dp059}, \cite{dp060}, \cite{dp482}\end{tabular} &
  \cite{dp688} &
  \cite{dp427}, \cite{dp442} \\ 
\rowcolor[gray]{0.96}
\begin{tabular}[c]{@{}l@{}}Performance \\ Enhancement\end{tabular} &
  \cite{dp019}, \cite{dp419}, \cite{dp033} &
  \begin{tabular}[c]{@{}l@{}}\cite{dp020}, \cite{dp666}, \\ \cite{dp709}, \cite{dp040}, \\ \cite{dp709},  \cite{dp160}, \\ \cite{dp707}\end{tabular} &
  \cite{dp427}, \cite{dp469} \\ 
\begin{tabular}[c]{@{}l@{}}Reduction of\\  Overhead\end{tabular} &
  \begin{tabular}[c]{@{}l@{}}\cite{dp140}, \cite{dp026}, \cite{dp016},\\  \cite{dp439}\end{tabular} &
  \cite{dp020} &
  \cite{dp427}, \cite{dp442} \\ 
\rowcolor[gray]{0.96}
\begin{tabular}[c]{@{}l@{}}Secure and \\ Reliable\end{tabular} &
  \cite{dp026}, \cite{dp444} &
  \cite{dp040} &
  \cite{dp469} \\ \hline
\end{tabular}
\end{table}

\subsection{Limitation of the Exiting Studies}
We have compiled the limitations of the current approaches and future research directions in the following section. 

\begin{itemize}

\item \textbf{Choosing appropriate value for  $\epsilon$:}
 One of the major concerns in differential privacy is to choose an appropriate value for privacy budget denoted with $\epsilon$. The value of $\epsilon$ determines the strictness and strength of privacy. A smaller value of $\epsilon$ provides stronger privacy, however, with that the data losses its utility and vice versa. Therefore, finding a optimal value for $\epsilon$ is a great challenge for any DP-based technique.There is very limited works have been done on finding the optimal value.
    
\item \textbf{Correlation of data:}
Real-world datasets often contain strong correlation among the data which can cause disclosure of an individual's information. For example, such correlation between data can enable an adversary to find out sensitive information about different individuals. The adversary can combine obfuscated data with existing correlation and derive sensitive information about individuals. Researchers have proposed model based approach   \cite{paper420_63},   \cite{paper420_101} and transformation based approach   \cite{dp261},   \cite{dp160} for solving this issue of data correlation. However, these approaches did not prove to be an optimal solution and even sometimes can distort the data to a great extent \cite{wang2017cts}. Therefore, overcoming the obstacles of data correlation is a big challenge for differential privacy.
    
\item \textbf{Sensitivity:}
The principal purpose of differential privacy is to maintain the indistinguishability between the presence or absence of any individual in the dataset. Sensitivity is the maximum difference between two neighboring dataset (datasets differing in one row). Noise is added to cover the difference and maintain the same identity for both the databases. To improve sensitivity more noise needs to be added. However, large value of noise can distort the data and this can result in unwanted utility loss. These trade-off between privacy and utility needs to be maintained. Some technologies are using diversity sensitivity to overcome this issue  \cite{paper420_92}. However, it is still a challenge to choose an optimal value of sensitivity and preserve both privacy and utility simultaneously to maintain the trade-off.

\item \textbf{Vulnerability of basic mechanisms:}
Basic mechanisms of differential privacy face various challenges when researchers tried to implement them. In  \cite{dp067}, authors have shown  Laplace Noise is vulnerable to tracker attack. After querying few times, results (after adding Laplace noise with true value) have either no privacy or no utility.  In addition, \cite{dp469} have shown Laplace mechanism is vulnerable to Data Reconstruction attack. 

\end{itemize}

Finally, we have analyzed the papers published until April 31, 2020. Between May to the acceptance time, new papers may have been published in different scientific journals. In addition, we have not considered papers that are not focused on wearable devices rather used traditional IoT devices or MIoT devices.

\subsection{Future Direction}


In this section, we discuss future research direction in differential privacy. 

\vspace{2mm}

\noindent\textbf{Adaptive Privacy Budgeting for real-time data:}  Selecting an appropriate $\epsilon$ value is a crucial task in order to protect the privacy of the individuals. Unlike static data, we do not have prior knowledge of the data point values for the real-time streaming data. Thus, distributing budget adaptively (rather than statically) could be an excellent way to preserve balance between privacy and utility. Kellaris et al's \cite{dp709} research work has established the superiority of adaptive budget allocation over static budget allocation for streaming data.

\vspace{2mm}
\noindent\textbf{Integrating Blockchain with Differential Privacy:} In the past few years, blockchain has emerged as a key technology that establishes trust among trust-less parties in a distributed and decentralized manner. It has the potential to transform the way in which we share information \cite{chowdhury2019comparative} and guarantees secure and immutable data storage.  Along with its association with the bitcoin concept, the blockchain has been widely adopted in various fields, including healthcare, finances, logistics, IoT \cite{xiong, banerjee,alnemari,chowdhury2020survey}, and even wearable devices and smart healthcare \cite{mettler}.However, privacy is a big concern for blockchain, specially for public blockchain. Several researchers have used differential privacy to overcome privacy issues in blockchain systems \cite{hassan_differential}. 



In addition, blockchain is also used to build trust on the privacy budget by providing a distributed and transparent system. Authors from \cite{zhao2021blockchain, han2020blockchain} have proposed a blockchain based approach for tracking and saving differential privacy costs. 




\vspace{2mm}
\noindent\textbf{Differential Privacy in Big data and Artificial Intelligence (AI):} In today's world big data and AI have become one of the major driving forces. In recent years, big data and artificial intelligence (AI) have gotten a lot of attention and have become valuable resources. Big data refers to the production of a massive amount of data from various sources, including sensors, wearable devices, IoT devices, social media platforms, and many more. Due to its massive scale, privacy and security are a major concern regarding this. Researchers are using differential privacy for big data publishing in different domains (e.g., transport, health)\cite{zhu2019}. Authors from \cite{shrivastva2014} have shown that integrating differential privacy have resulted in resolving many of the privacy issues of big data publishing. 


\section{Conclusion}
\label{conclusion}
Due to the rapid growth of wearable technologies, there is hardly any area which is not affected by it. Therefore, Health sectors are benefited by adapting wearable technologies. However, privacy is always a big concern for health data. Differential privacy has emerged as one of the most popular privacy preserving mechanisms in recent times. The purpose of this article is to understand the trends and limitations of differential privacy on wearable data.



 Even though the existing privacy-preserving mechanisms are applicable in wearable technologies, there is still a gap, which necessitates additional effort in designing and developing a more secure privacy preserving mechanism to hinder PII information. Though the proposed schemes from \cite{dp026, dp427, dp469} have outperformed existing state-of-the-art solutions by providing a more efficient solution  and have provided better data protection still these schemes suffer from overhead problems such as computational, communications, and even in-system overheads.

 There are still a number of issues faced by differential privacy such as fine-tuning $\epsilon$, its privacy budget, to balance between privacy and usability as well as other issues such as dimentionality and temporal correlation also need to be addressed. There are some open research problems such as data reconstruction errors \cite{dp160, dp469}, perturbation errors \cite{dp160}, absolute errors \cite{dp688} and relative errors \cite{dp026, dp040, dp707}. It is important to find solutions that can minimize an error rate significantly so that data utilization can be increased. In addition, we have also observed that there are only limited number of works have been done on real-time health data publishing.Finally, privacy mechanisms need to be more adaptive where users can fine-tune their privacy according to their needs.

\bibliographystyle{spphys}
\bibliography{main.bib}

\clearpage

\appendix

\section{Appendix: Paper Search and Review}
\label{ref:appendix}
\begin{figure*}[hb]
    \includegraphics[width= 0.8\linewidth, center]{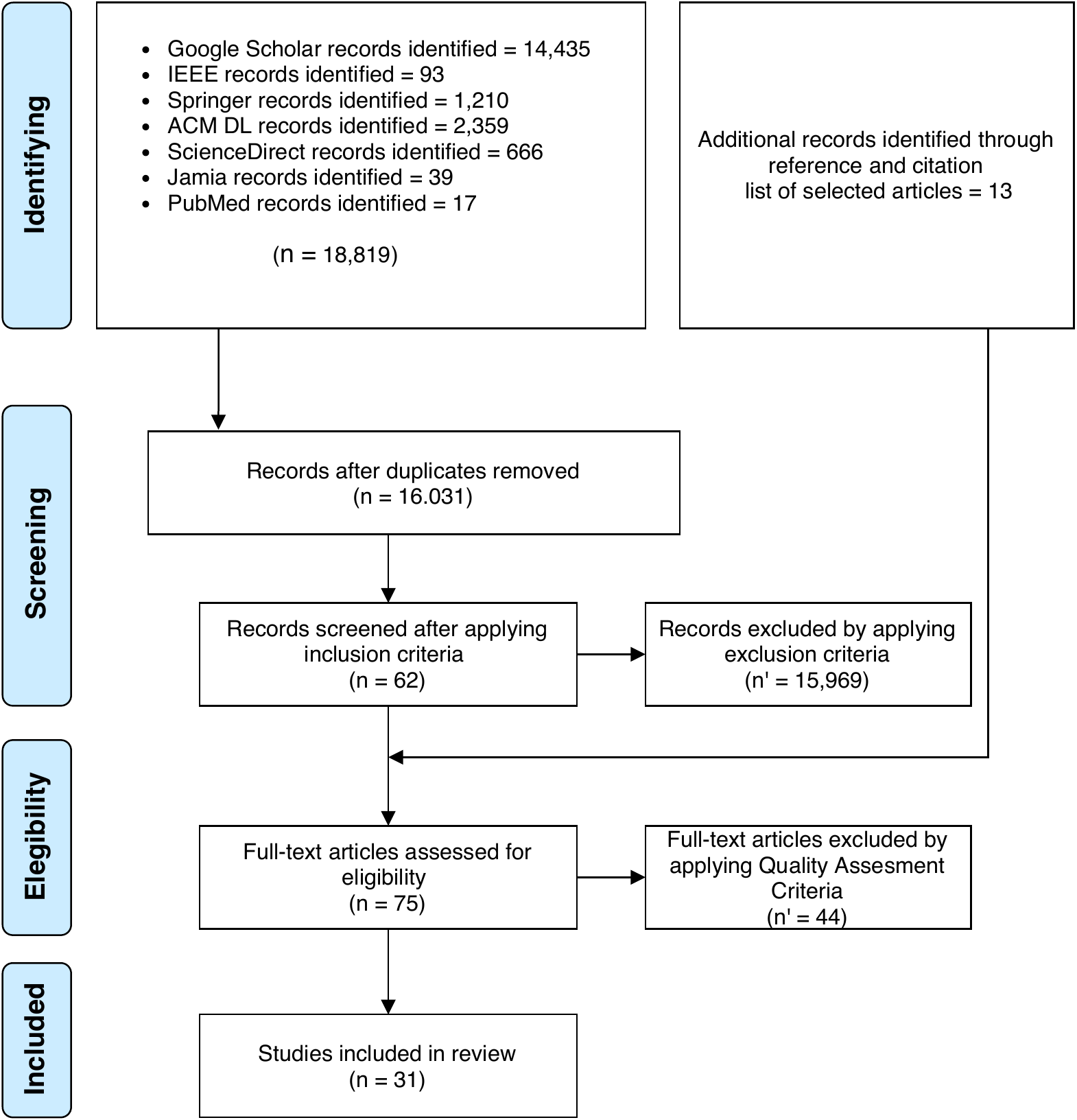}
    \caption{PRISMA flow diagram}
    \label{fig_prisma}
\end{figure*}

\begin{table*}[t]
\centering
\caption{Retrieved papers using logical AND \& OR }
\label{tab:my-table}
\resizebox{\textwidth}{!}{%
\begin{tabular}{l|ccccccc}
\hline
\cellcolor[gray]{0.9}
{ } &
  \multicolumn{7}{c}{\cellcolor[gray]{0.9}\textbf{Number of retrieved papers}} \\ \cline{2-8} 
\multirow{-2}{*}{{ \cellcolor[gray]{0.9}\textbf{Keywords (using logical AND \& OR)}}} &
  \multicolumn{1}{l|}{\cellcolor[gray]{0.9}\textbf{\begin{tabular}[c]{@{}l@{}}Google\\ Scholar\end{tabular}}} &
  \multicolumn{1}{l|}{\cellcolor[gray]{0.9}\textbf{IEEE}} &
  \multicolumn{1}{l|}{\cellcolor[gray]{0.9}\textbf{\begin{tabular}[c]{@{}l@{}}ACM\\ DL\end{tabular}}} &
  \multicolumn{1}{l|}{\cellcolor[gray]{0.9}\textbf{\begin{tabular}[c]{@{}l@{}}Science\\  Direct\end{tabular}}} &
  \multicolumn{1}{l|}{\cellcolor[gray]{0.9}\textbf{Springer}} &
  \multicolumn{1}{l|}{\cellcolor[gray]{0.9}\textbf{JAMIA}} &
  \multicolumn{1}{l}{\cellcolor[gray]{0.9}\textbf{PubMed}} \\ \hline \hline

\rowcolor[gray]{0.96}
\begin{tabular}[c]{@{}l@{}}(Wearable OR "privacy preserving") AND \\ ("data publishing" AND "differential privacy")\end{tabular} &
  3170 &
  45 &
  157 &
  132 &
  288 &
  0 &
  4 \\ \hline
\begin{tabular}[c]{@{}l@{}}("Temporal data" OR "wearable data") AND \\ ("publish using" OR "publish under") AND \\ ("differential privacy")\end{tabular} &
  0 &
  0 &
  0 &
  27 &
  2 &
  0 &
  0 \\ \hline
\rowcolor[gray]{0.96}
\begin{tabular}[c]{@{}l@{}}(Wearable OR Medical OR Health) AND \\ "data privacy" AND (using OR under) AND \\ "differential privacy"\end{tabular} &
  5920 &
  0 &
  311 &
  188 &
  472 &
  14 &
  3 \\ \hline
\begin{tabular}[c]{@{}l@{}}("Wearable" OR "Wearable devices generated") AND \\ "data privacy" AND (using OR under) AND \\ "differential privacy"\end{tabular} &
  901 &
  0 &
  63 &
  54 &
  66 &
  0 &
  0 \\ \hline
\rowcolor[gray]{0.96}
\begin{tabular}[c]{@{}l@{}}((Review OR Survey OR SLR OR “Literature Review”) on AND\\  ("Wearable Data" OR "Wearable devices data")) AND \\ (("Publishing using" OR "data publishing") AND \\ ("differential privacy")\end{tabular} &
  4 &
  0 &
  177 &
  1 &
  0 &
  0 &
  0 \\ \hline
\begin{tabular}[c]{@{}l@{}}(Wearable OR "privacy preserving") AND \\ (("data publishing" OR "publishing using") AND \\ "differential privacy")\end{tabular} &
  3170 &
  45 &
  157 &
  209 &
  289 &
  0 &
  10 \\ \hline
\rowcolor[gray]{0.96}
\begin{tabular}[c]{@{}l@{}}(Wearable data) AND (publish under OR publishing using) AND \\ "differential privacy"\end{tabular} &
  1270 &
  3 &
  1494 &
  55 &
  93 &
  25 &
  0 \\ \hline \hline
{ \textbf{Total retrieved:}} &
  { \textbf{14435}} &
  { \textbf{93}} &
  { \textbf{2359}} &
  { \textbf{666}} &
  { \textbf{1210}} &
  { \textbf{39}} &
  { \textbf{17}} \\ \hline
\end{tabular}%
}
\end{table*}



\end{document}